\documentclass[fleqn,usenatbib]{mnras}

\usepackage{newtxtext,newtxmath}

\usepackage[T1]{fontenc}
\usepackage{ae,aecompl}

\usepackage{graphicx}	
\usepackage{amsmath}	
\usepackage{amssymb}	
\usepackage[bottom, hang, flushmargin]{footmisc}
\usepackage{threeparttable}
\usepackage{xspace}
\usepackage{adjustbox}

\newcommand{\civ}{\,\ion{C}{iv}\xspace}

\newcommand{\Chandra}{\,\emph{Chandra}\xspace}

\newcommand*{\thead}[1]{\multicolumn{1}{c}{ #1}}

\title[Extreme quasar X-ray variability]{The frequency of extreme X-ray variability for radio-quiet quasars}

\author[J. D. Timlin et al.]{
John D. Timlin III,$^{1,2 \thanks{E-mail: jxt811@psu.edu}}$
W. N. Brandt,$^{1,2,3}$
S. Zhu,$^{1,2}$
H. Liu,$^{4,5}$
B. Luo$^{4,5}$
and Q. Ni$^{1,2}$
\\
$^{1}$Department of Astronomy \& Astrophysics, 525 Davey Lab, The Pennsylvania State University, University Park, PA 16802, USA \\
$^{2}$Institute for Gravitation and the Cosmos, The Pennsylvania State University, University Park, PA 16802, USA \\
$^{3}$Department of Physics, 104 Davey Lab, The Pennsylvania State University, University Park, PA 16802, USA\\
$^{4}$School of Astronomy and Space Science, Nanjing University, Nanjing, Jiangsu 210093, China\\
$^{5}$Key Laboratory of Modern Astronomy and Astrophysics (Nanjing University), Ministry of Education, Nanjing, Jiangsu 210093, China\\
}

\date{Accepted 2020 August 27. Received 2020 August 27; in original form 2020 July 14}

\pubyear{2020}

\begin{document}
\label{firstpage}
\pagerange{\pageref{firstpage}--\pageref{lastpage}}
\maketitle

\begin{abstract}
We analyze 1598 serendipitous \Chandra \hbox{X-ray} observations of 462 radio-quiet quasars to constrain the frequency of extreme amplitude \hbox{X-ray} variability that is intrinsic to the quasar corona and innermost accretion flow. The quasars in this investigation are all spectroscopically confirmed, optically bright ($m_i \leq 20.2$), and contain no identifiable broad absorption lines in their optical/ultraviolet spectra. This sample includes quasars spanning $z \approx$ 0.1 -- 4 and probes \hbox{X-ray} variability on timescales of up to $\approx 12$ rest-frame years. Variability amplitudes are computed between every epoch of observation for each quasar and are analyzed as a function of timescale and luminosity. The tail-heavy distributions of variability amplitudes at all timescales indicate that extreme \hbox{X-ray} variations are driven by an additional physical mechanism and not just typical random fluctuations of the coronal emission. Similarly, extreme \hbox{X-ray} variations of low-luminosity quasars seem to be driven by an additional physical mechanism, whereas high-luminosity quasars seem more consistent with random fluctuations. The amplitude at which an \hbox{X-ray} variability event can be considered extreme is quantified for different timescales and luminosities. Extreme \hbox{X-ray} variations occur more frequently at long timescales ($\Delta t \gtrsim 300$ days) than at shorter timescales, and in low-luminosity quasars compared to high-luminosity quasars over a similar timescale. A binomial analysis indicates that extreme intrinsic \hbox{X-ray} variations are rare, with a maximum occurrence rate of $<2.4\%$ of observations. Finally, we present \hbox{X-ray} variability and basic optical emission-line properties of three archival quasars that have been newly discovered to exhibit extreme \hbox{X-ray} variability.
 
 
 
\end{abstract}

\begin{keywords}
galaxies: active -- quasars: general -- quasars: supermassive black holes -- X-rays: general -- X-rays: galaxies
\end{keywords}



\section{Introduction}\label{sec:intro}

Encircling the super-massive black hole (SMBH) and inner accretion disk at the center of quasars and active galactic nuclei (AGN) is a ``corona'' of hot gas in which thermal UV photons are reprocessed into X-ray photons through Compton up-scattering (e.g. \citealt{Galeev1979, Haardt1991, Jiang2014}). Much of the emitted \hbox{X-ray} radiation generated by this process is directly observed as the power-law continuum in the quasar \hbox{X-ray} spectrum (e.g. \citealt{Mushotzky1993, Reynolds2003}). This intrinsic \hbox{X-ray} emission is found to be ubiquitous in quasars (e.g. \citealt{Gibson2008}; Pu et al. submitted); however, the basic nature and physical properties of this corona remain uncertain. Understanding the observable properties of the intrinsic \hbox{X-ray} emission in quasars will help to constrain further models of the corona and innermost accretion flow.

Variability of the \hbox{X-ray} emission from AGN is a useful probe of the underlying nature of the coronal region. For example, by measuring the characteristic timescales of \hbox{X-ray} variability, \citet{McHardy2006} demonstrated that the accretion process of the SMBH in AGN has similarities to that of smaller black holes. Moreover, the observed red-noise-like \hbox{X-ray} power spectrum in AGN resembles that of \hbox{X-ray} binaries, suggesting that a similar physical mechanism in these two classes of objects produces the \hbox{X-ray} photons (e.g. \citealt{Green1993, Uttley2002}). \hbox{X-ray} variability from AGN has been observed over a wide range of timescales from hours (e.g. \citealt{Ponti2012}) to years (e.g. \citealt{Vagnetti2011, Gibson2012, Shemmer2017}). These timescales have been used to understand better the properties of the corona. For example, the short-timescale variations indicate that the corona is centrally located near the SMBH (e.g. \citealt{Mushotzky1993}). 

The amplitude of \hbox{X-ray} variability is another useful characterization of the underlying physics in the corona. Large-scale investigations of quasars have found that the amplitude of \hbox{X-ray} variability intrinsic to the quasar corona and innermost accretion flow generally increases with increasing timescale, yet typically does not exceed a factor of $\approx 2$ (e.g. \citealt{Gibson2012}; \citealt{Middei2017}). In some rare cases, there have been AGN that have exhibited extreme \hbox{X-ray} variations, which are often defined as variations in the \hbox{X-ray} flux by a factor of ten or more. These large-amplitude fluctuations, however, are not always generated by variations in the coronal region.

External effects have sometimes been linked to extreme \hbox{X-ray} variability in different quasar populations. For example, \hbox{X-ray} variations in quasars that contain broad absorption lines (BALs) in their UV spectra have been linked to changes in the column density of the obscuring material along the observer's line-of-sight. In this case, absorption of the \hbox{X-ray} photons is responsible for the observed variability (e.g. \citealt{Gallagher2002, Gibson2009, Saez2012, Giustini2016}). The \hbox{X-ray} absorption in BAL quasars has been attributed to gas in the large-scale outflows as well as stalled ``shielding" gas closer to the SMBH. Additionally, \hbox{X-ray} variations of radio-loud quasars are often associated with variations of the jet-linked \hbox{X-ray} component instead of coronal variations (e.g. \citealt{Carnerero2017}). While these objects provide information about the nature of the quasar environment, they generally do not grant robust insight into the intrinsic variations of the quasar corona and innermost accretion flow.

There is a small subset of non-BAL, radio-quiet AGN that also exhibit extreme \hbox{X-ray} variations intrinsic to the corona and innermost accretion flow, many of which are identified as narrow-line Seyfert 1 galaxies with low-to-moderate luminosity and black-hole mass. One notable example is the low-$z$, changing-look AGN 1ES 1927+654 that was recently discovered to have undergone an extreme X-ray variation \citep{Ricci2020}. This variation was attributed to the destruction and re-creation of the inner accretion disk and corona, perhaps by interactions between the accretion disk and debris from a tidally disrupted star. Extreme X-ray variations in higher luminosity quasars are seemingly much more rare, where only eight such objects have been confirmed to exhibit extreme \hbox{X-ray} variations over the past twenty years (e.g. \citealt{Strotjohann2016, Liu2019, Ni2020}). A loose constraint on the rate of intrinsic extreme \hbox{X-ray} variability in quasars was briefly estimated as part the analysis of \citet{Gibson2012}, which investigated the general \hbox{X-ray} variability properties of spectroscopically-confirmed quasars; however, both the small sample size and the presence of \hbox{X-ray} upper limits in the sample affected the statistical power of their constraint. Two of the eight quasars were recently found to vary in \hbox{X-ray} flux by more than a factor of ten (SDSS J0751+2914, \citealt{Liu2019}; SDSS J1539+3954, \citealt{Ni2020}). Given the apparent rarity of these extreme events, it was somewhat surprising that these two events were found at nearly the same time. A larger-scale systematic investigation is therefore warranted in order to understand better the frequency of these extreme variations.

In this investigation, we aim to constrain better the frequency of extreme \hbox{X-ray} variability that is intrinsic to the quasar corona or due to other changes in the central accretion flow (hereafter we will refer to the combination of these two phenomena as intrinsic \hbox{X-ray} variability). Such intrinsic variability could be due to changes in the coronal emission, or due to \hbox{X-ray} absorption by a thick inner accretion disk as in the case for quasars with weak emission lines and quasars with high Eddington ratio (e.g. \citealt{Luo2015, Liu2019, Ni2020}). To constrain better this frequency, we assembled a large, unbiased sample of radio-quiet quasars that are devoid of BALs and have multiple, high-quality \hbox{X-ray} measurements. This large sample provides a sufficient number of quasars to more tightly constrain the frequency of intrinsic extreme \hbox{X-ray} variability. Studying the occurrence rate of extreme \hbox{X-ray} variation of the innermost accretion flow provides insight into the nature of the corona, and thus will help inform physical models of the \hbox{X-ray} emission from typical quasars in general.

This paper is organized as follows: Section 2 describes the data sets used to assemble a sample of typical quasars for this investigation, and presents the methods used to find \hbox{X-ray} counterparts. The data-analysis techniques used to reduce and analyze the \hbox{X-ray} data are discussed in Section 3. Section 4 presents the variation of quasars over time, presents the frequency of extreme \hbox{X-ray} variability among typical quasars, and discusses the implications of these results. Serendipitously discovered extremely \hbox{X-ray} variable quasars are presented in Section 5, and our results are summarized in Section 6. Throughout this work, we adopt a flat $\Lambda$-CDM cosmology with \mbox{$H_{0}$ = 70 km s$^{-1}$ Mpc$^{-1}$}, \mbox{$\Omega_M$ = 0.3}, and \mbox{$\Omega_{\Lambda}$ = 0.7}, and we utilize the \Chandra Interactive Analysis of Observations (CIAO; \citealt{Fruscione2006}) version 4.10\footnote{\url{http://cxc.harvard.edu/ciao/releasenotes/ciao_4.10_release.html}} software and {\mbox{CALDB version 4.8.3.\footnote{\url{http://cxc.harvard.edu/caldb/}} }}


\section{Sample Selection}\label{sec:sample_selection}

To assemble our quasar catalog, we combined the SDSS data release fourteen quasar catalog (DR14Q; \citealt{Paris2018}) and the large quasar catalog from \citet{Richards2015} which compiled other spectroscopically-confirmed quasars that overlap the SDSS imaging footprint. A brief description of these two survey catalogs is given below along with the method used to find serendipitous, multi-epoch \Chandra observations. We also describe the methods used to flag and remove quasars that have BALs or strong radio emission. 


\subsection{Optical data}

The SDSS DR14Q \citep{Paris2018} catalog compiled optical properties of all $526,356$ quasars that were observed in the first three SDSS projects (\citealt{York2000}; \citealt{Eisenstein2011}) as well as the first data release of the fourth SDSS project \citep{Dawson2016}. In total, DR14Q contains quasars over an area of $\approx$9376 deg$^2$ across the sky that span a  wide redshift range, the majority of which are between $0.1\leq z \leq4$. The redshifts reported in the catalog have been either visually measured or obtained using a principal component analysis (PCA) on the source spectrum (e.g.;\ \citealt{Paris2012}). Also included is a measurement of the balnicity index ({\tt{BI\_CIV}}; \citealt{Weymann1991}), which indicates the presence of a broad absorption line near the \civ emission line. Additionally, the quasars in the DR14Q catalog have been matched to the objects detected in the Faint Images of the Radio Sky at Twenty centimeters (FIRST; \citealt{Becker1995}) survey, and the radio fluxes have been recorded for matching quasars. Analysis of the frequency of extreme \hbox{X-ray} variability for typical quasars requires that the \hbox{X-ray} counterparts be almost always detected; therefore we also impose an empirically determined restriction on the apparent magnitude of $i\leq20.2$ because optically bright quasars tend also to be brighter in \hbox{X-ray}s. This sample contains $222,358$ total quasars.

To the DR14 quasar catalog, we added quasars from the comprehensive quasar catalog presented in  \citet{Richards2015}. This catalog mainly contains quasars from SDSS-I/II/III (\citealt{York2000}; \citealt{Eisenstein2011}) through data release 10 (DR10), and thus these quasars are reported in the DR14Q catalog; however, it also includes spectroscopically-confirmed quasars from the 2dF quasar redshift survey (2QZ; \citealt{Croom2004}), the 2SLAQ survey \citep{Croom2009}, and the AGES project \citep{Kochanek2012} that lie within the SDSS footprint but have no SDSS spectrum. For each quasar in this comprehensive catalog, \citet{Richards2015} reported the photometric information from SDSS, along with the redshift measurements from the respective survey, which allows us to straightforwardly combine these data with the information from the DR14Q catalog. Imposing the same $i$-band magnitude cut as before, this catalog returns an additional $21,669$ quasars not reported in the DR14Q catalog. Although the \citet{Richards2015} quasar catalog adds a relatively small number of quasars compared to the DR14Q sample, our scientific goal of constraining the frequency of extreme X-ray variability requires as many quasars as possible; therefore, we elected to retain these objects in our analysis.

\begin{figure}
	\includegraphics[width=\columnwidth]{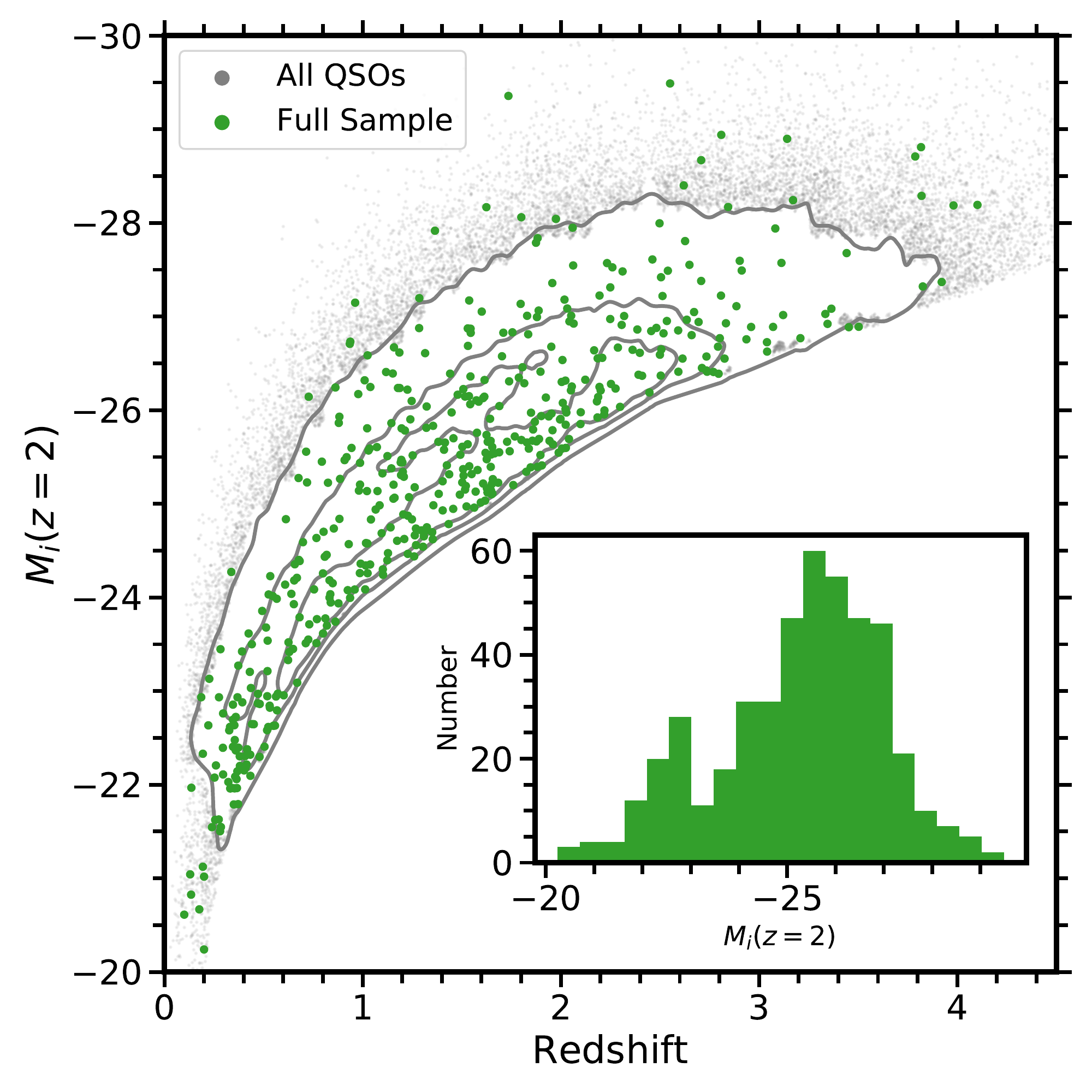}
    \caption{Absolute $i$-band magnitude (corrected to $z=2$; \citealt{Richards2006}) as a function of redshift for all of the confirmed quasars that were used to search for serendipitous \Chandra observations. The grey contours enclose 35, 68, and 95 percent of the quasars. The green points depict the quasars in our Full sample that have multiple, high-quality \Chandra \ \hbox{X-ray} observations as described in Section \ref{sec:typical_qsos}. The inset plot shows the $M_i$ distribution of the quasars in the Full sample.}
    \label{fig:Mi_z2}
\end{figure}

In total, we obtain $244,027$ bright quasars that we used to search the \Chandra database for serendipitous observations. The quasars assembled from the aforementioned catalogs form a reasonably homogeneous combination of typical, blue type-I quasars. We depict the absolute magnitude of the quasars in this sample as a function of their redshift in Figure \ref{fig:Mi_z2}. Even after imposing a restriction on the apparent brightness, this full quasar sample spans a wide range in luminosity and redshift.

\subsection{Finding \Chandra counterparts}

In this investigation, we elected to perform \hbox{X-ray} photometry of the SDSS quasars in the \Chandra events files rather than simply match these quasars to the \Chandra Source Catalog (CSC; \citealt{Evans2010}). Performing the analysis in this manner allows us to examine these quasars to a greater sensitivity than would be provided by a simple match to the blind-search CSC, and it enables us to generate optimal constraints on quasars that are not detected. We closely followed the methods described in \citet{Timlin2020} to search for serendipitous counterparts in the \Chandra data, which will be outlined below. Restricting our analysis to only the serendipitously observed sources removes any potential biases that might arise due to the exceptional properties of targeted quasars. In this work, we only consider \Chandra observations through MJD=58504 (21 January 2019) which includes 7122 \Chandra observations used in the analysis. Our scientific results could be heavily affected by the presence of \hbox{X-ray} non-detections in the sample (see Section \ref{sec:ctrt_flx}); therefore, we only considered observations from \Chandra as opposed to other observatories (e.g. {\emph{XMM-Newton}}) because its low point-source background levels make it well suited to detecting serendipitous objects. \Chandra has been taking high-quality data for $\approx 20$ years which enables us to probe quasar variability on timescales that span a significant fraction of a human lifetime.

As in \citet{Timlin2020}, we began searching for serendipitous counterparts using the Multi-Order Coverage (MOC\footnote{\url{http://cxc.cfa.harvard.edu/cda/cda_moc.html}}) map, which approximates the regions of the public \Chandra observation footprints, to quickly remove quasars in our large sample that are not covered by a \Chandra observation. Next, the CIAO \citep{Fruscione2006} tool {\tt{find\_chandra\_obsid}} was used to find the observation ID associated with the given position of a quasar; we searched only for detections with the ACIS instrument \citep{Garmire2003} where no gratings were used in the observation. We retained all observations of any given quasar regardless of how many times it was observed. Quasars were required to lie no less than 30 pixels ($\approx15\arcsec$) from the edge of a detector to remove non-detections and low-quality measurements due to the quasar lying partially outside of the detector.

\begin{figure}
	\includegraphics[width=\columnwidth]{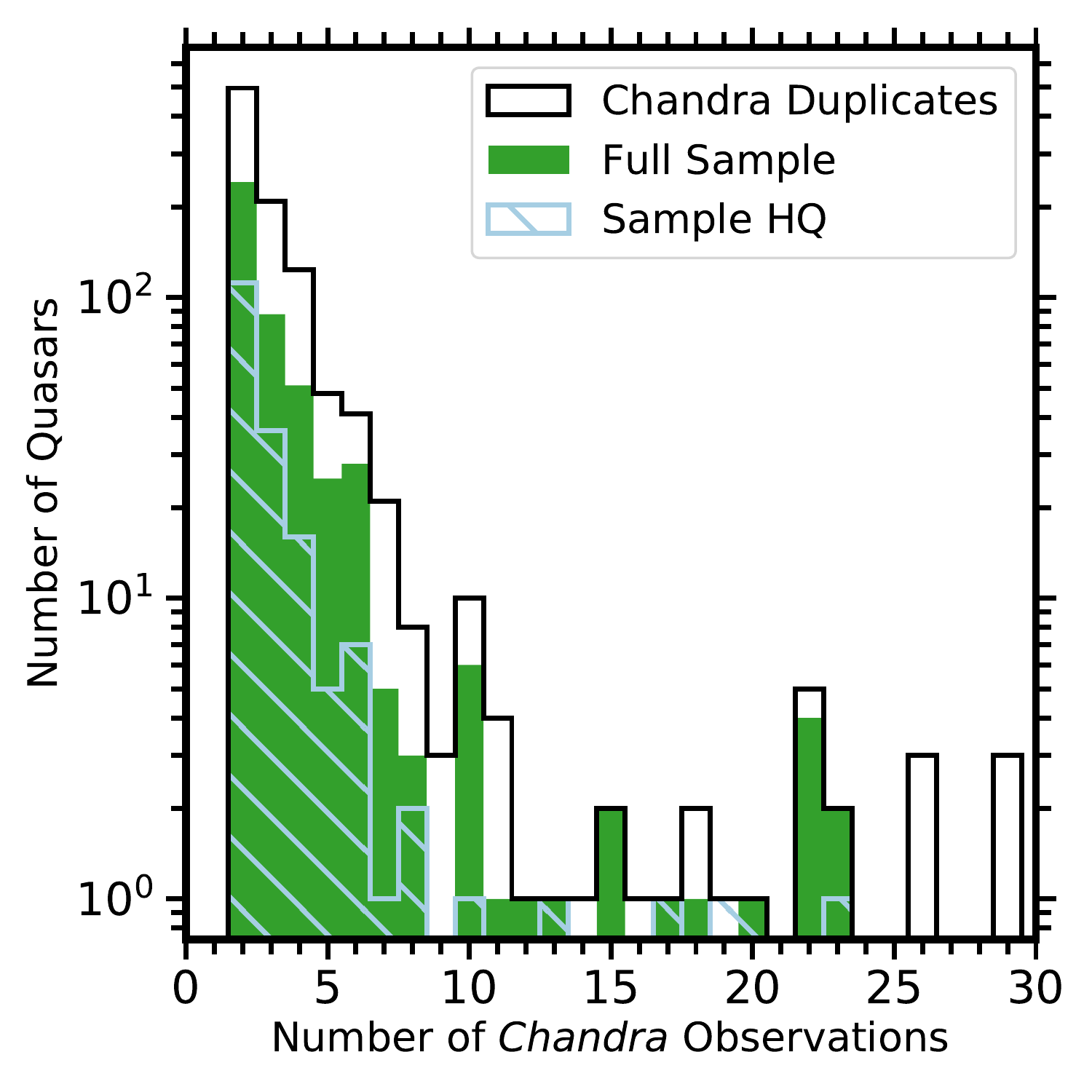}
    \caption{The number of serendipitous \Chandra observations that comprise the \hbox{X-ray} light-curve for each of the quasars in our samples. The open black histogram depicts the initial sample of $878$ quasars with multiple \Chandra observations that was assembled in Section \ref{sec:sample_selection}. The green and blue hatched histograms show the number of \Chandra observations of the quasars in our Full sample and Sample HQ (which has a 100\% \hbox{X-ray} detection fraction), respectively, as defined in Section \ref{sec:typical_qsos}. We find that most quasars have been serendipitously observed between two and five times with \Chandra, however there are cases where \Chandra has serendipitously observed the same quasar $\geq10$ times.}
    \label{fig:Nepoch}
\end{figure}

In total, we found $12,283$ \Chandra observations of $7,813$ quasars that passed the above criteria. For the purpose of this investigation, we were only interested in keeping quasars that had more than one observation in order to compute the variation between the epochs. After removing observations with only a single epoch of observation, $4,035$ observations of $1,160$ quasars remained. Quasars with off-axis angles $<0.5\arcmin$ were removed to ensure that the \hbox{X-ray} coverage was serendipitous. Additionally, since the point-spread function (PSF) size and vignetting effects increase at large off-axis angles, we also required that our sample lie within 9\arcmin \ of \Chandra pointing positions which restricted our sample to $3,155$ observations of $878$ quasars. We depict the distribution of the number of \Chandra observations per quasar for this sample of 878 quasars in Figure \ref{fig:Nepoch}. The majority of quasars that were found serendipitously have fewer than five \Chandra observations.


\section{Multi-wavelength Data Analysis}\label{sec:data_anaylsis}
\subsection{\hbox{X-ray} data reduction}\label{sec:Xdata_redux}
Each of the aforementioned observations was processed using standard CIAO tools \citep{Fruscione2006}, again closely following the method detailed in \citet{Timlin2020}. The primary and secondary data files were used to generate a bad pixel file, an event list, and a spectrum file using the {\tt{chandra\_repro}} tool. A background cleaning procedure was also implemented in this step and is dependent on the observation mode of the ACIS camera. Background flares were then removed using the {\tt{deflare}} tool, which flags the MJD of any spurious counts above the 3$\sigma$ level. 

From the flare-filtered, reprocessed events files, a full-band (0.5--7 keV), a soft-band (0.5--2 keV), and a hard-band (2--7 keV) image were created using the {\tt{dmcopy}} tool (using ASCA grades 0, 2, 3, 4, 6) for each observation. Sources were extracted from each of the three images using the {\tt{wavdetect}} tool with a false-detection probability of $10^{-6}$. The \hbox{X-ray} positions of the detected sources in each image were combined to form a complete detection list and the optically-determined position of the quasar was matched to this list. If the quasar position matched within $2\arcsec$ of the \hbox{X-ray} position of a source in the list, the \hbox{X-ray} position from {\tt{wavdetect}} was adopted; otherwise photometry was performed on the \hbox{X-ray} image at the optically-determined position of the quasar.

Counts were extracted at the source position using a matched-filter region instead of the circular regions used in \citet{Timlin2020}. While circular regions are a simple and effective aperture to use for positions near the aim-point, as the off-axis angle increases the size of the circle must also increase to surround the majority of the counts, which leads to a larger number of background counts being enclosed by the source region. The matched-filter technique instead used the best-fit elliptical region of a simulated source at the same position on the ACIS chip as the quasar. This method removes the excess background enclosed by a circular region, thus improving the sensitivity of the source detection.

Sources were simulated using the Model of AXAF Response to \hbox{X-ray}s (MARX; \citealt{Davis2012}) software suite. MARX is a ray tracing program that is designed to simulate realistic observations with the \Chandra telescope considering the true properties of an observation (e.g. MJD, detector ID, nominal pointing information, source location, the aspect solution file, and a real or simulated spectrum). MARX was run for each of the three energy-dependent images that were created for every observation. The MARX suite creates an image of the simulated photons similar to the observation data files for a user-defined exposure time. The {\tt{dmellipse}} tool was used to fit an elliptical region that enclosed 90\% of the simulated photons and was subsequently used to extract counts from the observation. All of the regions were visually inspected to ensure that the fitting procedure succeeded in describing the simulated images and that the regions were sensible to use for extracting counts in the observed image.

Background counts were extracted using circular annuli with inner (outer) radii of 15 (50) arcsec  surrounding the source position. If there is a source detected by {\tt{wavdetect}} that overlaps this region, we subtract the contribution of the elliptical region provided by {\tt{wavdetect}} from the total background. Additionally, we remove the sections of the background region that fall off of the chip boundary, forming a `pie'-shaped region (e.g. Pu et al. submitted). This method systematically defined the background regions that best described the local background near the source. All of these background regions were visually inspected. In a small number of cases when the quasar had many neighbors, we placed a circular background region in a nearby, source-free location for a more appropriate measurement of the background.

Exposure maps were created using the {\tt{flux\_image}} tool to quantify the effective exposure at the off-axis position of the quasar (the exposure maps generated in this work have units of photons$^{-1}$ cm$^{2}$ s). Additionally, the {\tt{flux\_image}} tool incorporates the decline of the ACIS quantum efficiency over time into each exposure map. Effective exposure time, which is the exposure time corrected for this loss of sensitivity, was computed by multiplying the exposure time of the observation by the ratio of the median value from the exposure map in the source region to the maximum value of the exposure map. This ratio represents the vignetting effects at the source position compared to that of an on-axis observation. We use the median value of the exposure map in the source region to reduce the effect of individual bad pixels within the source region. Exposure maps were created for each band using effective energies of 1 keV, 3 keV, and 2 keV for the soft, hard and full band, respectively.\footnote{We adopted effective energies similar to those presented in the {\tt{flux\_image}} documentation (\url{https://cxc.cfa.harvard.edu/ciao/ahelp/fluximage.html}). Reasonable changes in the energy were found to have little effect on the results.} The exposure maps were input into the {\tt{dmextract}} routine when extracting counts files from each of the three energy bands.

The source detection significance for each band was computed using the binomial no-source probability (e.g. \citealt{Broos2007}; \citealt{Xue2011}; \citealt{Luo2015}), $P_B$, using the formula:
\begin{equation}
P_B(X \geq S) = \sum^{N}_{X=S} \frac{N!}{X!(N - X)!}p^{X}(1 - p)^{(N - X)},
\end{equation}
where $S$ is the total source counts, $N$ is the total number of raw source and background counts, and $p = 1/(1+ f_{\rm area})$ where $f_{\rm area}$ is the ratio of the background to source region area. Since we are performing forced-photometry at the pre-specified positions of optically bright objects, we consider a source to be detected if $P_B \leq 0.02$ (2.3$\sigma$), and calculate the 1$\sigma$ errors on the net source counts following the numerical method of \citet{Lyons1991} and the method described in \citet{Gehrels1986}. We find that the vast majority ($\approx 96\%$) of the quasars in this investigation have a \Chandra detection significance of $P_B\ll 0.0001$, and thus relaxing the detection threshold to $P_B \leq 0.02$ does not introduce a large number of false positives into our analysis (see Section \ref{sec:typical_qsos} for more details). When a source is not detected we use the Poisson confidence interval method \citep{Kraft1991} to estimate a 90\% confidence upper limit on the counts. Finally, in this investigation we do not compute physical flux values from the counts, but rather we elect to work with count fluxes to eliminate additional uncertainties that come from estimating physical flux (see Section \ref{sec:ctrt_flx}).

\subsection{Radio-loud and BAL quasars}\label{sec:RL_BAL}

The goal of this investigation is to determine the frequency of intrinsic extreme \hbox{X-ray} variability among the majority population of radio-quiet quasars. Radio-loud quasars have traditionally been thought to exhibit \hbox{X-ray} emission associated with their jets in addition to coronal emission (e.g. \citealt{Miller2011, Zhu2020}). We also exclude BAL quasars from our sample since their observed \hbox{X-ray} emission can vary due to absorption changes in the central region instead of a change in coronal emission (See Section \ref{sec:intro}). In this investigation we flag and remove these two populations in the same manner as in \citet{Timlin2020} which we will briefly describe below.

To remove radio-loud quasars from our sample, we compute the radio-loudness parameter, $R = f_{6 \rm{cm}}/f_{2500}$, where $f_{6 \rm{cm}}$ and $f_{2500}$ are the quasar fluxes at rest-frame $6\ \rm{cm}$ and 2500 \AA , respectively \citep{Kellerman1989}. We convert the apparent $i$-band magnitude into the 2500 \AA\ flux following the method in Section 5 of \citet{Richards2006}. The rest-frame $6\ \rm{cm}$ flux is computed using the $20\ \rm{cm}$ flux (assuming a radio spectral index of $\alpha_{\nu}=-0.5$) from the Faint Images of the Radio Sky at Twenty-Centimeters (FIRST; \citealt{Becker1995}) survey, or the NRAO VLA Sky Survey (NVSS; \citealt{Condon1998}) database if the quasar is not covered by FIRST. For quasars that are not detected at $20\ \rm{cm}$, the $3\sigma$ upper limit on the flux was estimated as $0.25 + 3\sigma_{\rm rms}$ mJy, where $\sigma_{\rm rms}$ is the RMS flux at the source position and 0.25 mJy is the CLEAN bias correction for the FIRST survey (\citealt{White1997}; the CLEAN bias correction for NVSS is 0.3). 

Typically, a quasar is taken to be radio loud if $R > 10$ (e.g. \citealt{Kellerman1989}); however, previous investigations have found that significant \hbox{X-ray} contributions from the jet are not generally present until $R\gtrsim100$ (e.g. \citealt{Miller2011, Zhu2020}). We find that $\approx 43\%$ of our sample has $3\sigma$ upper limits of $R \leq 10$ whereas $\approx 94\%$ of the sample has upper limits of $R \leq 30$, which is considerably lower than the $R$-level at which the jet generally contributes significant \hbox{X-ray} emission. The quasars in this sample with $R \leq 30$ can therefore realistically be considered either radio-quiet or mildly radio-intermediate. In either case, the coronal \hbox{X-ray} emission should be the dominant source of \hbox{X-ray}s in these quasars; therefore we retain quasars in our sample with $R \leq 30$. In total, 32 radio-loud objects ($R > 30$) are flagged and removed from the sample.

The balnicity index can be used for a first-cut removal of BAL quasars in the sample (e.g. removing {\tt{BI\_CIV}}>0); however this parameter is only provided for the quasars in the DR14Q catalog. To identify and remove BALs in our full sample, we first fit the spectrum of each quasar using the PyQSOFit\footnote{\url{https://github.com/legolason/PyQSOFit}} \citep{PyQSOFit} software. Following the method described in Section 3.2 of \citet{Timlin2020}, we fit the global continuum of each quasar with a power-law model and the \ion{C}{iv} and \ion{Mg}{ii} emission lines with three Gaussian profiles when they were present in the quasar spectrum. Absorption troughs were identified as the $3\sigma$ flux outliers between the data and the best-fit model, and their width was measured as the distance between the two pixel locations where the trough intersected with the model. When the width was $\geq$ 2000 $\rm{km\ s^{-1}}$, we flagged the object as a BAL quasar. At low-$z$ ($z\leq1.7$) the \ion{C}{iv} emission line has shifted out of the UV spectrum, so we instead identified BALs to the lesser extent possible in the \ion{Mg}{ii} emission-line region since quasars with \ion{Mg}{ii} BALs almost always harbor \ion{C}{iv} BALs (e.g. \citealt{Zhang2010}). A lack of a \ion{Mg}{ii} BAL does not necessarily demonstrate that the spectrum is devoid of BALs; however, we retain such quasars in our sample. Using the fraction of identified BALs at high-redshift and the \ion{Mg}{ii} BALs already discovered at low-redshift in our sample, we conservatively estimate that unidentified BAL quasars will comprise $\lesssim 10\%$ of the Full sample (defined in the next Section). The number of unidentified BAL quasars that remain in our sample should be sufficiently small as to not have a large impact on the results. In total, 44 BAL quasars were robustly identified and removed from the sample.

\subsection{The typical quasar sample}\label{sec:typical_qsos}

\begin{figure}
	\includegraphics[width=\columnwidth]{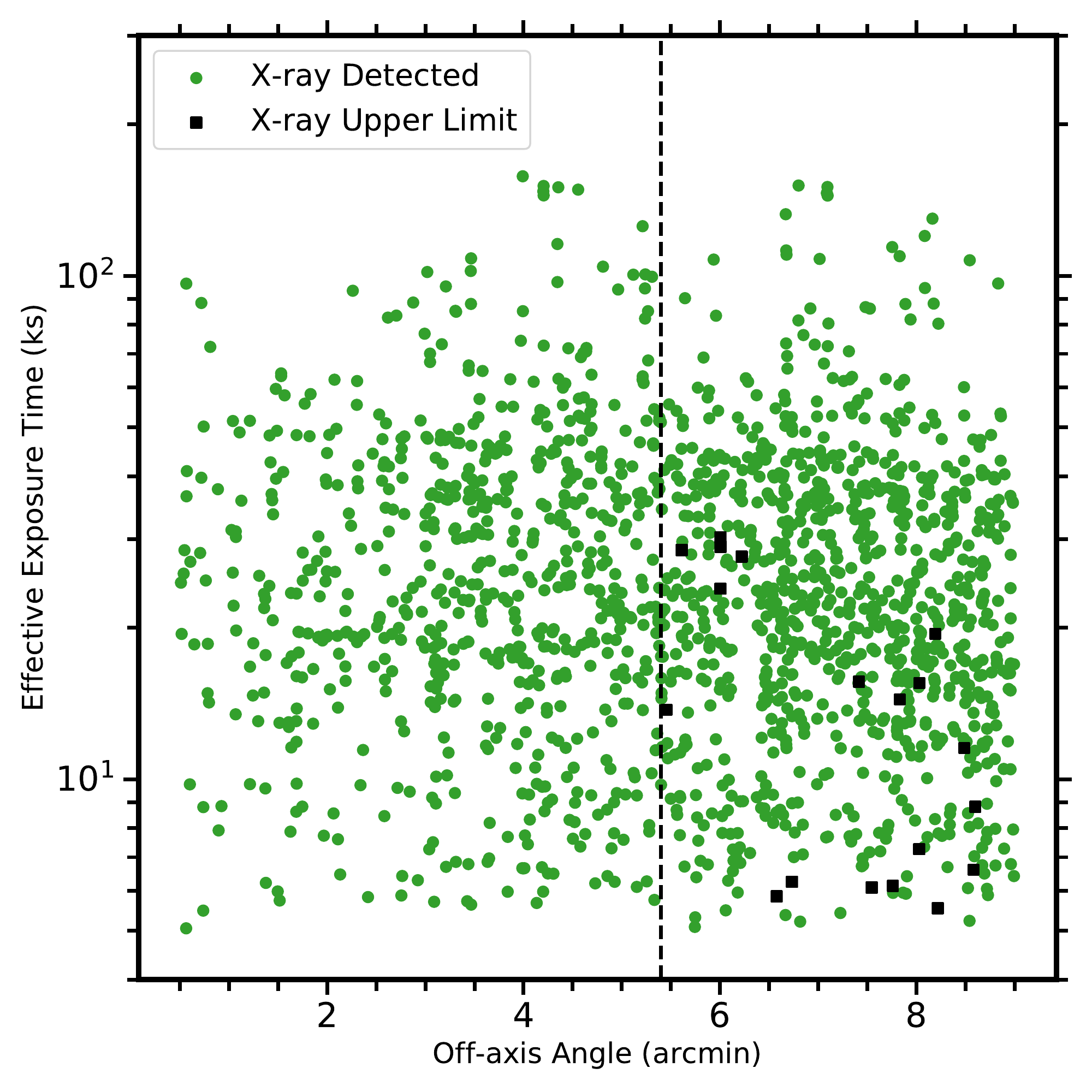}
    \caption{Effective exposure time as a function of off-axis angle for each of the $1,598$ \Chandra observations in our Full sample. Green points depict observations that are \hbox{X-ray} detected in at least one band according to the binomial no-source probability, whereas black squares represent quasars that are not detected in any \hbox{X-ray} band. The vertical dashed black line shows the location of the cut to generate Sample HQ. All of the quasars with off-axis angle less that threshold ($5.4\arcmin$) are \hbox{X-ray} detected, and thus the analysis of these objects in Section \ref{sec:freq_EXV} will not be affected by \hbox{X-ray} upper limits.}
    \label{fig:Texp_theta}
\end{figure}

In addition to the initial restrictions on the data ($i$-band magnitude, \Chandra MJD, and off-axis angle) as well as the removal of BAL and radio-loud quasars, some final cuts were made after the \hbox{X-ray} images were processed. First, quasars that overlap an ACIS chip gap were removed from the sample since the sensitivity in these regions is much lower than for the rest of the chip and can result in low-sensitivity non-detections. We also removed twenty quasars that lie near \hbox{X-ray} bright clusters of galaxies since the clusters can significantly and non-uniformly increase the background level in the source and background regions which lowers the sensitivity of the observation. Furthermore, we removed observations that are less sensitive by requiring that all of our quasars have an effective exposure time of at least 5 ks. Figure \ref{fig:Texp_theta} depicts the effective exposure time as a function of off-axis angle for the quasars considered in the work. For the small background levels in a single \Chandra observation and off-axis angles considered in this investigation, a flat cut in effective exposure time roughly corresponds to a cut in the limiting flux of an observation. At off-axis angles larger than $\approx9\arcmin$, the PSF size and vignetting effects become very large making the observations less sensitive. 

After imposing these further criteria, $1,598$ high-quality observations of $462$ quasars remained in the sample (hereafter the Full sample; see Table \ref{tab:Samples}). We find that only 19 of the $1,598$ \Chandra observations (1.1\%) had quasars that were \hbox{X-ray} undetected and thus have an upper limit on the \hbox{X-ray} counts. In this investigation, we consider an object to be \hbox{X-ray} detected if the binomial no source probability is $ P_B \leq 0.02$ in any band. As mentioned in Section~\ref{sec:Xdata_redux}, the majority of the observations in the Full sample are detected with a high significance ($P_B \ll 0.0001$). Since the measurements are independent of each other, we can sum the no-source probabilities to estimate the number of false positives present in the sample. The total number of false-positive sources expected in this sample is $ \approx 0.7$ using $ P_B \leq 0.02$ as the detection threshold.


Any \hbox{X-ray} upper limits in our sample could lower the statistical power of our final result; therefore, we also create a high-quality sample which has a 100\% \hbox{X-ray} detection fraction by reducing the cut in off-axis angle from $9\arcmin$ to $5.4\arcmin$ (dashed vertical line in Figure~\ref{fig:Texp_theta}). This high-quality sample (hereafter, Sample HQ; see Table~\ref{tab:Samples}) contains $583$ \Chandra observations of $185$ quasars and will not be affected by upper limits. Both the Full sample and Sample HQ are dominated by quasars that have less than five serendipitous \Chandra observations as shown in Figure \ref{fig:Nepoch}. The Full sample of quasars is available online in machine-readable format (see Appendix \ref{append:A} for more details).

\begin{table}
\centering
\caption{Serendipitous quasar sample properties}
\label{tab:Samples}

\begin{tabular}{l l r l r l c}
\hline
Sample & $N_{\rm Q}$ & $N_{\rm obs}$ & $f_{\rm detected} $ &  $N_{\rm pairs}$ & $ \langle m_i \rangle $  &  $ \langle M_i(z=2) \rangle $\\
\thead{(1)} & \thead{(2)} & \thead{(3)} & \thead{(4)} & \thead{(5)} & \thead{(6)} & \thead{(7)}\\
\hline
Full & 462 & 1598 & 0.988 & 2567 & 19.2 & $-$25.54  \\
Full* & 462 & 1144 & 0.987 & 899 & 19.2 & $-$25.54  \\
HQ & 185 & 583 & 1.000 & 818 & 19.3 & $-$25.56 \\
HQ* & 185 & 443 & 1.000 & 331 & 19.3 & $-$25.56 \\
\hline
\end{tabular}

\begin{flushleft}
\footnotesize{{\it Notes:} Basic properties of the different samples used throughout this paper. Each row provides the information for the samples considered in this work, described in Sections \ref{sec:sample_selection} and \ref{sec:typical_qsos} (the ``*'' indicates that the \hbox{X-ray} light-curves have been down-sampled to three epochs using the method in Section \ref{sec:ctrt_flx}). Columns 2 and 3 provide the total number of quasars and \Chandra observations, and column 4 records the detection fraction of the observations in column 3. Column 5 presents the number of epoch pairs in the sample (see Section \ref{sec:ctrt_flx}), and columns 6--7 report the average values of the $i$-band apparent magnitude and absolute magnitude of each sample.}
\end{flushleft}

\end{table}


\section{The frequency of Extreme \hbox{X-ray} variability}\label{sec:freq_EXV}

\subsection{The count flux ratio}\label{sec:ctrt_flx}

We perform our analyses in terms of count flux \mbox{(cts cm$^{-2}$ s$^{-1}$)} rather than physical flux (erg cm$^{-2}$ s$^{-1}$) to remove additional uncertainty that comes from fitting a spectral model to the \hbox{X-ray} data. The count flux is defined as the ratio between the background-subtracted counts of each source and the average effective area at the quasar position in the exposure map (the exposure maps incorporate the off-axis effective area and quantum efficiency of the observation, and are in units of photons$^{-1}$ cm$^{2}$ s). In this investigation, we use the counts in the observed-frame full band (and the respective exposure map) to compute the count flux.\footnote{Quasars in nine observations were not detected in the full band but instead were detected in the soft (seven quasars) or hard (two quasars) bands. In these cases, the measured counts from the full band were used to compute the count flux. All nine observations have a binomial no-source probability in the full band that is only slightly larger than the detection threshold (with a maximum value of $P_B \approx 0.04$), and thus performing forced photometry at the known source location provides a reasonable estimate of the the full-band counts.} Uncertainties in the count flux are computed by propagating through the count errors from Section~\ref{sec:Xdata_redux}. 

In this investigation, we measure the amplitude of \hbox{X-ray} variability in the {\emph{observed}}-frame full band (0.5--7 keV), where the ACIS instrument is most sensitive, instead of converting to a common rest-frame bandpass. Choosing an ideal rest-frame bandpass well suited to studying all of the quasars in the sample would be challenging since the sample spans a wide range in redshift. Creating such a rest-frame bandpass would require that, for quasars at different redshifts, the observed-frame energy range be reduced and counts be extrapolated from this smaller bandpass. This would reduce the sensitivity of the measurements by removing energies where \Chandra is most effective for typical observations and greatly increase statistical uncertainties. Using the measurements in the observed-frame full band ensures that each observation is performed with the maximum sensitivity provided by \Chandra which is critical in this work since our goal of constraining the frequency of extreme \hbox{X-ray} variability requires that the sample have a very high detection fraction. \hbox{X-ray} non-detections may occur for several reasons, one of which being that the quasar has varied by an extreme amplitude. Using \hbox{X-ray} limits to compute the amplitude of variability, which is defined in this work as the ratio of count fluxes between two time-ordered epochs (see below), yields either an upper or lower limit on the amplitude. Incorporating both upper and lower limits (left and right censored data) into some statistical tests and analysis techniques is not straightforward, and improper treatment of the limits (e.g. inappropriate assumptions about their underlying distribution) might substantially impact the outliers of the distribution. The extreme \hbox{X-ray} variability events are the outliers in the variability amplitude distribution; therefore, it was imperative that the most sensitive \Chandra observations were used in order to reduce the number of \hbox{X-ray} limits in the sample.

Since the observed-frame bandpass probes different rest-frame energies depending on the quasar redshift, combining the variability amplitudes of different quasars presumes that different parts of the quasar \hbox{X-ray} spectrum vary in a basically similar manner. Previous investigations have found that the \hbox{X-ray} spectral slope is not highly dependent on redshift (e.g. \citealt{Just2007, Green2009}), which implies that a similar power-law trend should be present at the rest-frame energies that are being observed. Furthermore, investigations of individual Seyfert 1 AGN have demonstrated that the \hbox{X-ray} flux variations between the hard and soft bands are dominated by a change in normalization of the spectrum as opposed to different fluctuations in the bands; although, a secondary constant component was also found at hard \hbox{X-ray} energies, which has been associated with Compton reflection (e.g. \citealt{Taylor2003}, \citealt{Vaughan2004}). \citet{Gibson2012} found evidence for this two component model in quasars as well. Using a large, heterogeneous sample of \hbox{X-ray} detected quasars, \citet{Serafinelli2017} found a ``softer when brighter" trend which may similarly indicate the existence of this two component model of \hbox{X-ray} variability in quasars. Their investigation also found no clear evidence that this trend depends on redshift, \hbox{X-ray} luminosity, black-hole mass, or Eddington ratio. These previous investigations demonstrated that the \hbox{X-ray} variability of quasars is dominated by a constant normalization over the \hbox{X-ray} spectrum, in agreement with the assumption in this work; however, the secondary constant component suggests that this assumption may not be perfect, and thus further investigation is required to determine the magnitude of this effect on high-energy \hbox{X-ray} variability. Such an investigation is outside of the scope of our work. Therefore, the results reported below should therefore be interpreted as the variability of each quasar given the available \hbox{X-ray} data in the observed-frame full band.

The ratio of full-band count fluxes between \Chandra epochs was used to quantify the amplitude of \hbox{X-ray} variability. In order to space evenly this fractional change in flux around a ratio of one (e.g. no change in flux), we used $\rm{log}_{10}(\rm{count\ flux\ ratio})$ as our metric for variability. The timescale of this variability is defined as the difference between the start times (converted to the rest-frame) of the two \Chandra observations. As shown in Figure~\ref{fig:Nepoch}, many of the quasars in the Full sample and Sample HQ have only two epochs of \Chandra observations, and thus have only one measurement of variability and timescale. For quasars with more than two observations, the count flux ratio was measured between every unique pair of epochs which yields $N(N-1)/2$ permutations of variability measurements and timescales, where $N$ is the number of times the quasar has been observed with \Chandra. The  number of pairs for each sample is provided in Table \ref{tab:Samples}.

Quasars in the Full sample and Sample HQ with $>10$ serendipitous \Chandra observations were down-sampled such that there are only ten observations in the \hbox{X-ray} light-curve. This down-sampling reduced the effect of a single quasar with a large number of epoch permutations on the overall distribution of the count flux ratios (hereafter, we refer to these as the ``All permutations'' samples). A simple unbiased down-sampling method was employed that retained only the earliest observation, the latest observation, and eight randomly selected epochs between the two. In addition to reducing the effect that any one quasar had on the sample, this method also maintained the large temporal separation between \hbox{X-ray} observations of each quasar. Even after this down-sampling, these samples could still be biased by the heavy weighting of some multiply observed objects. To reduce further the effect that any single quasar had on the investigation, we again down-sampled the \Chandra light-curves of each quasar to three epochs using the same method. The count flux ratios were also computed for these two additional samples (hereafter referred to as the ``down-sampled permutations'' samples; see Table \ref{tab:Samples}).

\begin{table*}
\centering
\caption{Results of splitting the Full sample with reduced permutations by timescale and luminosity}
\label{tab:stats}

\begin{tabular}{l l r r r r r r r r}
\hline
 \thead{Sub-Sample} & & \multicolumn{2}{c}{Normality Test} & \multicolumn{2}{c}{Symmetry Test} & \multicolumn{2}{c}{Kurtosis}  &  \multicolumn{2}{c}{Median-based Statistics}  \\
\cline{3-4} \cline{5-6}\cline{7-8} \cline{9-10}
 \thead{Name} & \thead{$N_{\rm pairs}$} &  \thead{$A^2$} & \thead{$p$-value} & \thead{$A^2$} & \thead{$p$-value} & \thead{$k$} & \thead{$\sigma_k^{\rm a}$} & \thead{Median} & \thead{MAD}\\
 
\thead{(1)} & \thead{(2)} & \thead{(3)} & \thead{(4)} & \thead{(5)} & \thead{(6)}& \thead{(7)} & \thead{(8)} & \thead{(9)} & \thead{(10)}  \\
\hline
\multicolumn{9}{c}{Split by Timescale}\\
\hline
Short & 284 &  2.26 & $9.6\times10^{-6}$& 0.37 & 0.24 & 1.604 & 0.442 & 0.007 & 0.068  \\
Intermediate & 297  &  1.72 & $1.9\times10^{-4}$ & $-$0.57  & 0.63 & 1.570 & 0.759 &  $-$0.002 & 0.110  \\
Long & 318  &  2.21 & $1.3\times10^{-5}$ & $-$0.48 & 0.58 & 3.520 & 1.184 &  0.053 & 0.134 \\
\hline
\multicolumn{9}{c}{Split by $L_{2500}$}\\
\hline
Low & 380  &  5.08 & $1.4\times10^{-12}$ & $-$0.85 & 0.86 & 5.106 & 1.740 &  $-$0.001 & 0.102 \\
High & 278  &  1.58 & $4.7\times10^{-4}$ & 0.03 & 0.34 & 0.780 & 0.362 &  0.035 & 0.103 \\
\hline
\hline

\hline
\end{tabular}
\begin{flushleft}
\footnotesize{{\it Notes:}
Results of the statistical tests performed on the sub-samples in Section \ref{sec:dt} and \ref{sec:lum} for the Full sample with reduced epoch permutations. Column 1 and column~2 present the sub-sample name and the number of epoch pairs in that sub-sample, respectively. Column 3 reports the critical value of the AD normality test. We present the $p$-value of this test in column 4. Columns 5--6 are similar, but report the results of the AD two-sample test to determine if the distributions are symmetric. The kurtosis value and the error derived using a bootstrap re-sampling method are reported in columns 7--8. Columns 9--10 report the median and median-absolute-deviation (MAD) of the sample. While we only report the statistics from this sample, we found that the results from the other three samples yield similar conclusions. \\
$^{\rm a}$The standard deviation of the distribution of kurtosis values from 5000 bootstrap re-sampling iterations. }
\end{flushleft}

\end{table*}

\subsection{Extreme \hbox{X-ray} variability and timescale}\label{sec:dt}

\begin{figure*}
	\includegraphics[width=0.9\textwidth]{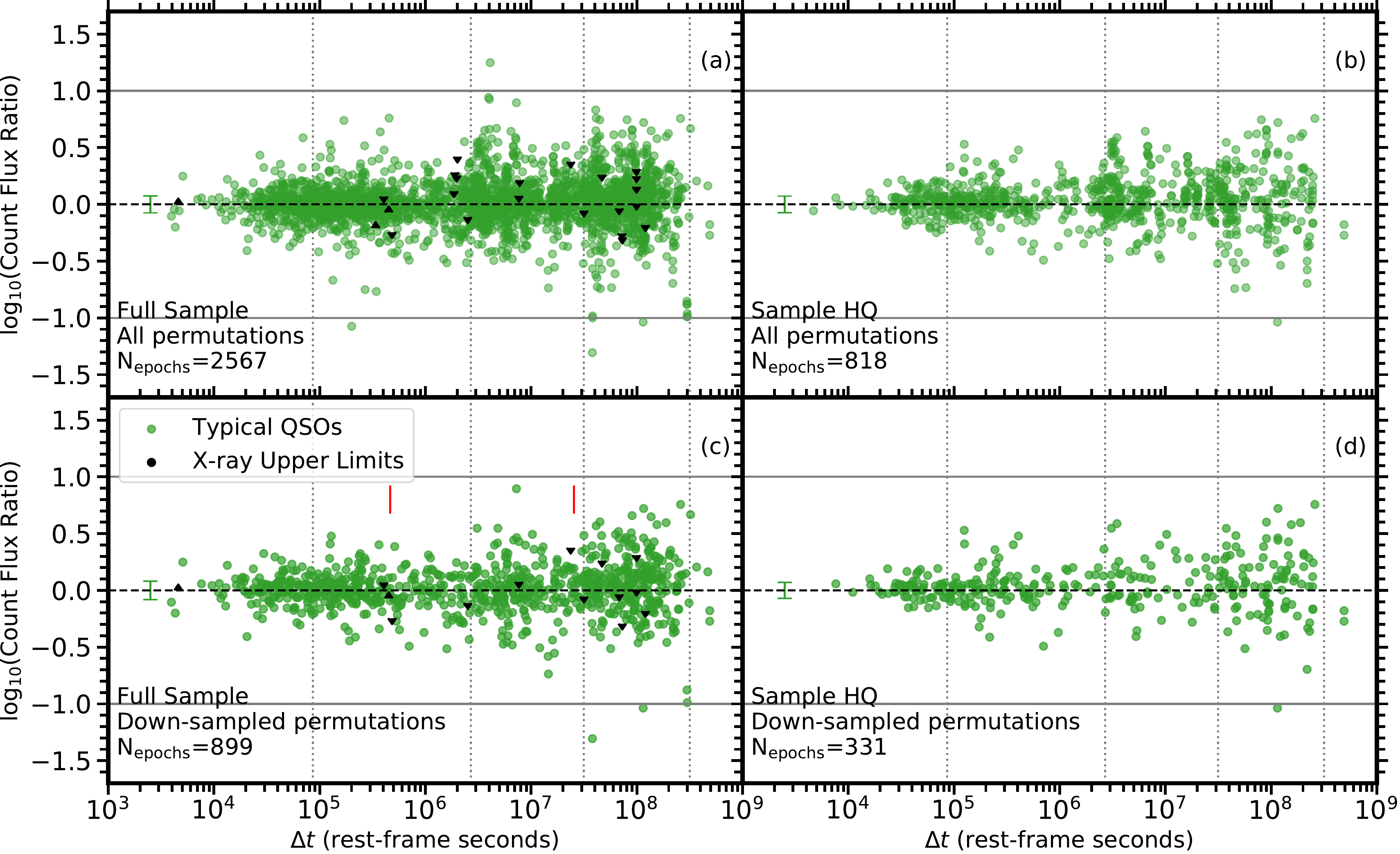}
    \caption{Count flux ratio as a function of the rest-frame time difference between two \Chandra epochs for the quasars in each sample. Panels (a) and (b) depict this ratio for every epoch pair for all of the quasars in the Full sample and Sample HQ, respectively. We mitigate the epoch-permutation effect in panels (c) and (d) for the Full sample and Sample HQ, respectively, by only considering the earliest and latest observations along with a randomly drawn epoch in the middle. Green data points depict epoch pairs where both measurements are \hbox{X-ray} detected, and black arrows represent when one of the measurements in a pair is an upper limit. The direction of the arrow indicates whether the \hbox{X-ray} limit is in the numerator (down arrow) or the denominator (up arrow) of the ratio. The black dashed line shows when the count flux ratio is unity, and the two grey lines show a count flux ratio of ten. The vertical dotted lines  in each panel correspond to timescales of one day, one month, one year, and ten years from left to right, respectively. The error bars represent the median error of the count flux ratio in each sample and the short red vertical lines in panel (c) mark where we split the data into three timescale bins (see Section \ref{sec:dt}). These panels show that \hbox{X-ray} variability by a factor of ten or more is extremely rare in quasars.  }
   \label{fig:Ctrt_all}
\end{figure*}

Previous investigations of \hbox{X-ray} variability in quasars have found that the \hbox{X-ray} emission can vary on timescales of days to years (e.g. \citealt{Gibson2012, Vagnetti2016, Shemmer2017}). In this work, we specifically seek to investigate the behavior of extreme \hbox{X-ray} variations over time. We therefore examined the count flux ratio as a function of timescale, $\Delta t$, in Figure \ref{fig:Ctrt_all} for the epoch pairs in the Full sample (panel (a); 2567 pairs) and in Sample HQ (panel (b); 818 pairs), as well as the down-sampled Full sample (panel (c); 899 pairs) and Sample HQ (panel (d); 331 pairs). The green points depict the count flux ratio of pairs where both observations provide \hbox{X-ray} detections. When one observation is an \hbox{X-ray} upper limit, we use the the 90\% confidence value from Section \ref{sec:Xdata_redux} to determine a limit on the count flux ratio (black arrows in Figure \ref{fig:Ctrt_all}). Arrows pointed down (up) represent the \hbox{X-ray} limit being in the numerator (denominator) in the ratio. Three epoch pairs were removed from the sample because neither observation provided an \hbox{X-ray} detection. Uncertainty in the count flux ratio was computed by propagating the errors in the counts using the method from \citet{Lyons1991} (green error bars show the median uncertainty in the data set). Figure \ref{fig:Ctrt_all} clearly shows that typical quasars do not generally vary by more than a factor of $\approx3$ $(\rm{log_{10}(count\ flux\ ratio)} \approx 0.5)$ in any of our samples, and that larger amplitude variations are rare.

\begin{figure}
	\includegraphics[width=\columnwidth]{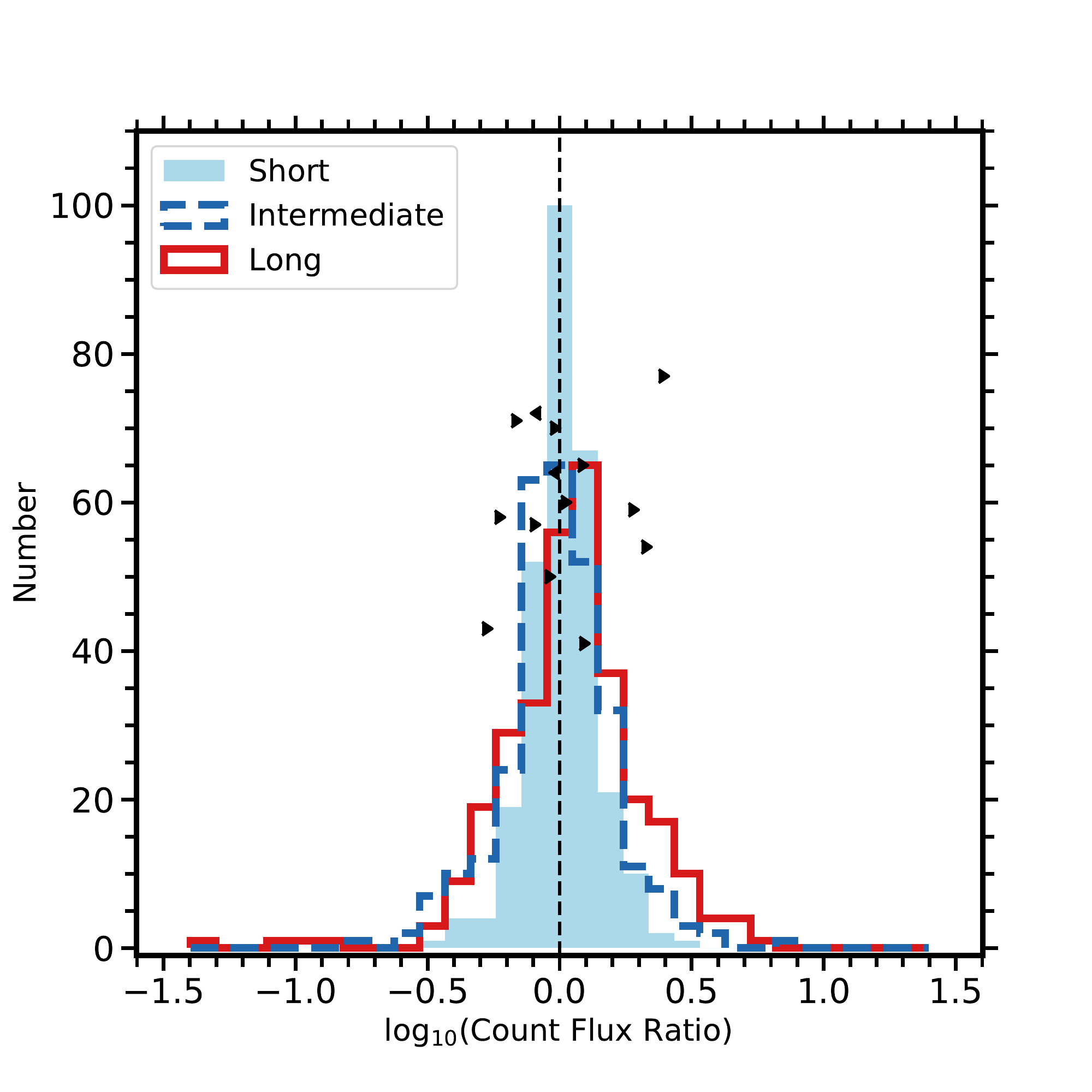}
    \caption{Distributions of count flux ratio for observations that are separated by $\Delta t\leq 0.463\times10^6\ \rm{s}$ (filled light blue, 284 points), $0.463\times10^6 < \Delta t \leq 25.4\times10^6\ \rm{s}$ (dashed blue; 297 points), and $\Delta t > 25.4\times10^6\ \rm{s}$ (open red; 318 points) in the Full sample with down-sampled permutations. The vertical dashed line depicts a flux ratio of one, and the black arrows show the \hbox{X-ray} limits and directions. An AD test of normality indicates that none of the three distributions is consistent with a Gaussian distribution. Furthermore, the three distributions are not consistent with each other according to an AD two-sample test (short-intermediate $p$-value=1.6$\times 10^{-3}$; intermediate-long $p$-value=1.9$\times 10^{-3}$; short-long $p$-value=1.0$\times 10^{-5}$). A visual inspection of the three distributions suggests that the intermediate and long timescale bins display larger count flux ratios than the short timescale distribution which implies that extreme \hbox{X-ray} variations occur more frequently at larger $\Delta t$.}
   \label{fig:Ctflx_dt}
\end{figure}

Figure \ref{fig:Ctrt_all} also suggests that the variability factor generally becomes larger at longer timescales in each of the samples as suggested by previous investigations (e.g. \citealt{Gibson2012}). It is therefore more instructive to split the sample by timescale and investigate the extreme \hbox{X-ray} variability properties of these sub-populations. We split the samples into short ($\Delta t\leq 0.463\ \times10^{6}\ \rm{s}$; 284 measurements), intermediate ($0.463 < \Delta t \leq 25.4\ \times10^{6}\ \rm{s}$; 297 measurements), and long ($\Delta\rm{t} > 25.4\ \times10^{6}\ \rm{s}$; 318 measurements) timescale bins of approximately equal sample size as presented in Figure~\ref{fig:Ctflx_dt}. The short, intermediate, and long timescale bins roughly correspond to timescales of $\Delta t \lesssim 5.3\ \rm days$, $5.3 \lesssim \Delta t \lesssim294\ \rm days$, and $\Delta t \gtrsim 294\ \rm days$, with median values of 1.1 days, 61 days, and 3 years, respectively. We only depict the count flux distribution of the Full quasar sample with the down-sampled permutations (panel (c) from Figure \ref{fig:Ctrt_all}) here since it maximizes the number of quasars in the sample but minimizes the effect any one quasar has on the final results.\footnote{While we only report the results from this data set, we tested each of the samples and find similar conclusions.} Analyzing the shape and symmetry of these three distributions, as well as their similarity to each other, will provide insight into the nature of the extreme \hbox{X-ray} variations at short, intermediate, and long timescales. 

Previous analyses have found that the \hbox{X-ray} flux distribution of AGN can be generally well modeled by a log-normal distribution (e.g. \citealt{Uttley2005}). The product or ratio of two log-normal distributions results in a log-normal distribution, much like the sum or difference of two normal distributions returns a normal distribution. Therefore, the count flux ratio should follow a log-normal distribution and log$_{10}$(count flux ratio) should be Gaussian distributed. Extreme \hbox{X-ray} variability should be a rare phenomenon, and, much like extreme outliers in Gaussian distributions, can be attributed to random fluctuations in the physical mechanism that is producing the \hbox{X-ray} flux. If, however, there are additional physical processes that occasionally drive larger \hbox{X-ray} flux variations, an excess of objects should appear in the outlier tails of the log$_{10}$(count flux ratio) distribution compared to a Gaussian distribution. Conversely, if a mechanism exists that suppresses large variations in the \hbox{X-ray} flux, the log$_{10}$(count flux ratio) distribution would exhibit a large peak and smaller tails than expected from a Gaussian distribution. 

Before analyzing the three timescale distributions in detail, we first determined whether the dispersion of each distribution is intrinsic or if it is largely due to measurement error. The likelihood method in \citet{Maccacaro1988} was used to deconvolve the parent distribution from the distribution of measurement errors, assuming the parent distribution is consistent with being a Gaussian. The intrinsic dispersion of the short-timescale distribution is $\langle\sigma_{\rm{p}}\rangle = 0.069 \pm 0.007$ which suggests, when compared to the standard deviation of the distribution ($\sigma_{\rm{short}} = 0.132$), that the measurement errors provide a somewhat larger contribution to the spread of the log$_{10}$(count flux ratio) distribution at short timescales than the parent distribution. The intrinsic dispersions of the parent population for the intermediate and long timescale distributions are $\langle\sigma_{\rm{p}}\rangle = 0.172 \pm 0.014$ and $\langle\sigma_{\rm{p}}\rangle = 0.199 \pm 0.017$, respectively. Since these values are only somewhat smaller than the standard deviation of the intermediate ($\sigma_{\rm{intermediate}} = 0.207$) and long ($\sigma_{\rm{long}} = 0.258$) timescale sub-samples, the dispersion of log$_{10}$(count flux ratio) is largely intrinsic in both distributions. Recall, however, that this analysis assumes that the parent distributions are Gaussian. If they are non-Gaussian distributions (e.g. due to additional physical mechanisms that drive extreme variations) the relationship between the intrinsic scatter and dispersion due to the measurement errors becomes more difficult to model.

We tested the consistency of each of the three distributions with a Gaussian distribution using an Anderson-Darling (AD) test of normality \citep{ADtest}. We elected to use this test (as opposed to the more frequently used Kolmogorov-Smirnov test) since it is more sensitive to differences in the tails of the distributions which, in this work, are occupied by the large-amplitude variations. While this test does not account for censored data, the fraction of censored points is only $\approx1.5\%$ and was found to have little effect on the results of the test.\footnote{We tested the effect that the limits had on the AD normality test by drawing ten thousand random values from a uniform distribution to use as the limit values in the AD test. For upper limits on count flux ratio, the bounds of random drawing were set between the 90\% confidence limit value and the minimum of the full sample. For lower limits on the ratio, the bound was set between the 90\% confidence limit value and the maximum of the full sample. The results of the AD test with the limits included were consistent with the test without the limits.} All three of the timescale bins are inconsistent with a Gaussian distribution at a high confidence level (see Table \ref{tab:stats}). Within the constraints of our data, this implies that additional physical processes are present at all timescales making the distributions non-Gaussian. 


We also tested the symmetry of each distribution by performing an AD two-sample test \citep{ADtest_2samp} on each half of the data, split at log$_{10}$(count flux ratio) $=0$. In all three of the timescale sub-samples, the two halves of the distributions are consistent and thus we consider them to be symmetric. The symmetry of the distributions suggests that the variability seen in our sample occurs relatively equally in both the positive and negative directions. Physical mechanisms that cause a sudden rise followed by a gradual decrease in \hbox{X-ray} flux (akin to what is seen in supernovae light-curves) can be ruled out for the quasar population.

A visual inspection of Figure \ref{fig:Ctflx_dt} suggests that the intermediate and long timescale distributions display an excess of large \hbox{X-ray} variations compared to the short timescale distribution. An AD two-sample test was used to determine the consistency of the three distributions with each other. None of the distributions is consistent with each other according to this test (short-intermediate {\mbox{$p$-value}}=1.6$\times 10^{-3}$; intermediate-long $p$-value=1.9$\times 10^{-3}$; short-long $p$-value=1.0$\times 10^{-5}$); however, the largest inconsistency occurs between the short and long timescale distributions. Such a result may be expected since the mechanism driving this variability is able to sample a wider range of \hbox{X-ray} fluxes at long timescales compared to short timescales.

A measurement of the kurtosis of each sub-sample was also used to determine whether the amplitude of \hbox{X-ray} variability under-populates the tail (e.g. a lack of large variations) or over-populates the tail (e.g. an excess of large variations) of their respective distributions compared to a normal distribution. The kurtosis is defined as the ratio between the fourth central moment and the standard deviation to the fourth power of a distribution and largely indicates whether there is an excess of outlier data in the distribution compared to expectations for a Gaussian distribution (e.g. \citealt{Livesey2007, Westfall2014}). The kurtosis can therefore be used as a measure of the  frequency at which large \hbox{X-ray} flux variations are produced by the underlying physical mechanisms compared to a Gaussian random fluctuation. All three distributions display a positive kurtosis (see Table \ref{tab:stats}; $k=0$ indicates a Gaussian kurtosis) which confirms that they indeed are non-Gaussian and contain an excess of large \hbox{X-ray} flux variations. Extreme variations in these timescale distributions are therefore likely a consequence of additional physical mechanisms as opposed to the result of random fluctuations in coronal emission. 

Since the three distributions of $\rm{log_{10}(count\ flux\ ratio)}$ are symmetric and display an excess of kurtosis compared to a Gaussian distribution, we elected to measure the spread in the distributions using the median-absolute-deviation (MAD; \citealt{Maronna2006}), which measures the median value of the absolute deviations of each value in a distribution from the sample median. The MAD statistic is much less sensitive to the outliers of a distribution and therefore better estimates the standard deviation for non-Gaussian distributions. Furthermore, for a Gaussian distribution, MAD can be related to the standard deviation through the equation: {\mbox{$\sigma_{\rm MAD} = 1.483\times$ MAD}}. For the short, intermediate, and long timescale sub-samples, we find deviations of MAD $=0.068$, MAD $=0.110$, and MAD $=0.134$, respectively. In this work, we define an ``extreme'' variation as a 5$\sigma_{\rm MAD}$ ($7.415\times$MAD) deviation in the distribution. Using the measured MAD values in this definition, we determine that extreme \hbox{X-ray} variations occur when the count flux changes by a factor of $\approx 3.19$, $\approx 6.54$, and $\approx 9.85$ for the short, intermediate, and long timescales, respectively. Previous investigations have used a factor of ten to denote an extreme \hbox{X-ray} variation, which we find to be a reasonable value at long timescales. This quantification of extreme variations allows us to put some past discoveries in better context. For example, the factor of $> 20$ variation of the luminous quasar SDSS J1539+3954 on long timescales \citep{Ni2020} corresponds to a 6.5$\sigma_{\rm MAD}$ event. 

\subsection{Extreme \hbox{X-ray} variability and $L_{2500}$}\label{sec:lum}

The quasar sample in this investigation spans a wide range of $L_{2500}$ and thus can be used to investigate differences in extreme \hbox{X-ray} variability between quasars with bright and faint ultraviolet (UV) luminosity. To investigate the extreme \hbox{X-ray} variations of quasars at different luminosities, the Full quasar sample with reduced epoch permutations was split at the median luminosity (${\rm{log_{10}}}(L_{2500})=30.4\ \rm{erg}\ \rm{s}^{-1}$) to generate two sub-samples with similar sizes. The lower luminosity quasars, however, tend to span longer rest-frame timescales since they are preferentially detected at low redshift (see Figure~\ref{fig:Mi_z2}). Since a significant difference was found in \hbox{X-ray} variability between short and long timescales in Section \ref{sec:dt}, an appropriate comparison of the luminosity bins can only be made when the timescale distributions of the two luminosity sub-samples are similar. To do so, we located the upper and lower 10\% of timescales in the low-luminosity sub-sample ($\Delta t>10^{8.1}$ and $\Delta t<10^5$ s, respectively) and removed observations with timescales beyond those values from the low- and high-luminosity sub-samples. This approach removes the timescale outliers in both luminosity sub-samples, and thus mitigates the biases in timescale. An AD two-sample test reports that the $\Delta t$ distributions of the two luminosity sub-samples are similar ($p$-value $=0.494$) and thus can be appropriately compared.

\begin{figure}
	\includegraphics[width=\columnwidth]{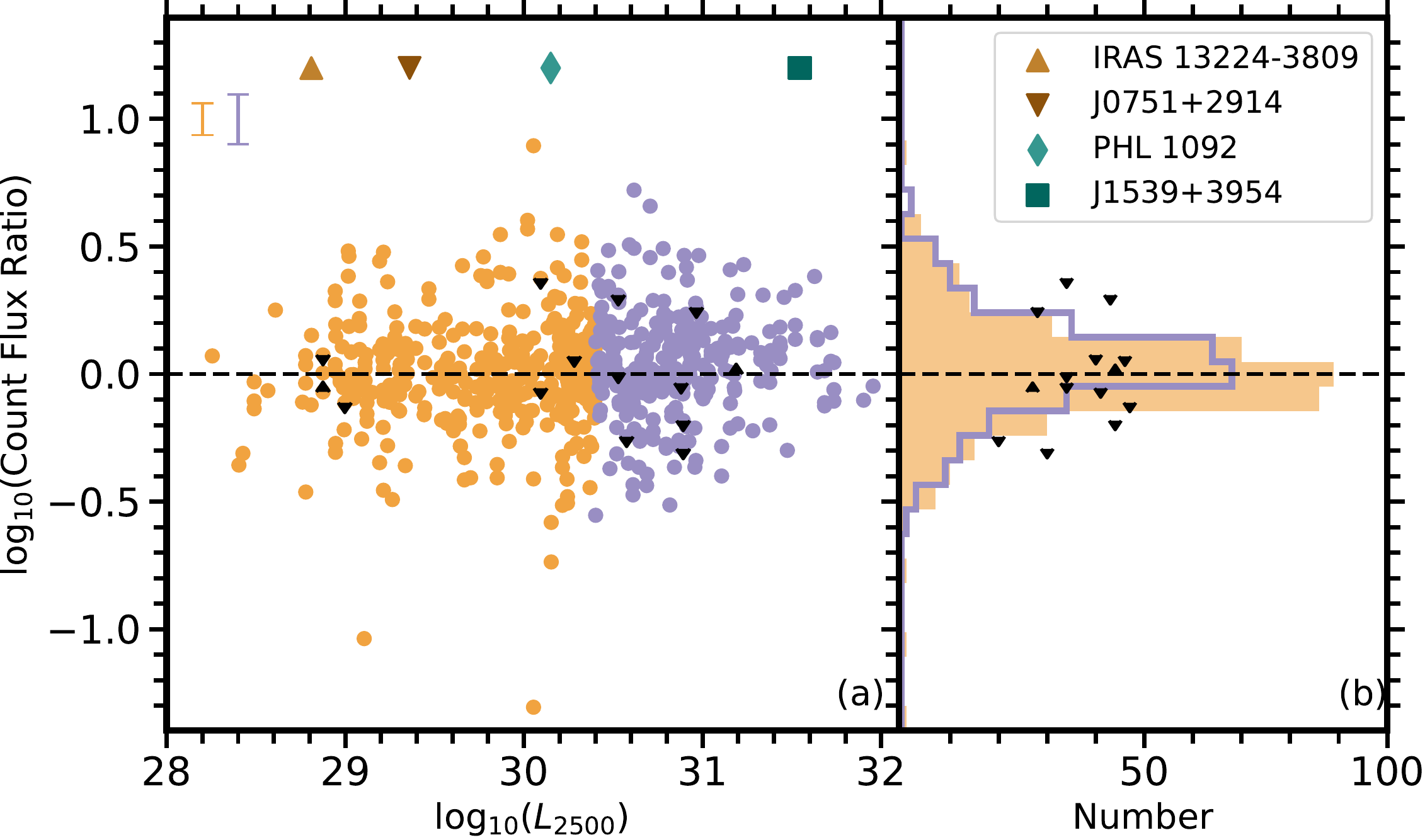}
    \caption{Panel (a): The logarithm of the count flux ratio as a function of the 2500 \AA \ luminosity for the Full sample of quasars with down-sampled epoch permutations. The data are split by the median value of $L_{2500}$ into bright (purple) and faint (orange) sub-samples. \hbox{X-ray} limits are shown by the black points, the horizontal dashed line corresponds to a count flux ratio of one, and the median error bars for each sub-sample are depicted in the top left corner. Also depicted are four examples of extremely \hbox{X-ray} variable AGNs in the literature to show with which luminosity bin they are associated (their log$_{10}$(count flux ratios) are forced to $1.2$). Panel (b): The distributions of the log$_{10}$(count flux ratio) in the two luminosity bins. While these two distributions appear to be similar, we find that the distribution of the low-luminosity quasars is not consistent with a Gaussian with high confidence, whereas the high-luminosity quasar distribution is much more Gaussian-like (though still non-Gaussian). The low-luminosity quasar distribution is also much more tail-heavy (as determined by the kurtosis) than the high-luminosity quasars. This suggests that the low-luminosity quasars are more likely to exhibit extreme \hbox{X-ray} variability than would be expected from random fluctuations. }
   \label{fig:Ctflx_Lum}
\end{figure}

Panel (a) of Figure \ref{fig:Ctflx_Lum} depicts log$_{10}$(count flux ratio) as a function of $L_{2500}$ for both the bright (purple points; 278 measurements of 165 quasars) and faint (orange points; 380 measurements of 201 quasars) sub-samples, and the histogram of their log$_{10}$(count flux ratio) is depicted in panel (b). As before, the intrinsic dispersion of both the low-luminosity and high-luminosity sub-samples were deconvolved from the distribution of measurement errors using the method from \citet{Maccacaro1988}. The standard deviation of the low-luminosity sub-sample and the high-luminosity sub-sample ($\sigma_{\rm{low}} = 0.218$ and $\sigma_{\rm{high}} = 0.200$, respectively) are only somewhat larger than the dispersion of their parent populations ($\sigma_{\rm{p, low}} = 0.177 \pm 0.015$ and $\sigma_{\rm{p, high}} = 0.150 \pm 0.011$, respectively) which implies that the observed scatter is largely intrinsic. 

In panel (a) of Figure \ref{fig:Ctflx_Lum}, we also depict the extremely \hbox{X-ray} variable AGN PHL 1092 (e.g.\ \citealt{Miniutti2012}), IRAS 13224$-$3809 (e.g.\ \citealt{Buisson2017}), SDSS J0751+2914 \citep{Liu2019}, and SDSS J1539+3954 \citep{Ni2020}, all of which have varied by more than a factor of $\approx20$ and span a wide range of $L_{2500}$ (we force the count flux ratio to be $\approx1.2$ for each of these objects to maintain the scale of our sample). Three of these four quasars have luminosities that overlap with the low-luminosity sub-sample defined in this investigation, while only SDSS J1539+3954 is consistent with the high-luminosity sub-sample. The luminosity distribution of these previously discovered extremely \hbox{X-ray} variable quasars anecdotally suggests that the lower luminosity quasars are more likely to exhibit extreme variations than the high-luminosity objects.


As before, the consistency of the luminosity distributions with a normal distribution was tested using an AD test of normality. The AD test reported that the low-luminosity sample is not consistent with a Gaussian with a high significance ($p$-value $= 1.4 \times 10^{-12}$), whereas the high-luminosity sample is more consistent with a Gaussian distribution ($p$-value $= 0.00047$). The AD test of the high-luminosity sub-sample has a much larger $p$-value and thus the null hypothesis of consistency with a Gaussian cannot be rejected with high confidence. This may suggest that the mechanism driving the extreme \hbox{X-ray} variations in luminous, typical quasars without strong radio emission or BALs may be significantly weaker than at low-luminosities. If, however, the timescale of this physical mechanism is dependent on physical properties of the quasar (e.g. black-hole mass or luminosity), then the extreme \hbox{X-ray} variations in the high-luminosity quasars may be just as prevalent as in the low-luminosity sample, but may occur on an even longer timescale than what is probed by this analysis.

A test of the similarity between the two count flux ratio distributions depicted in panel (b) of Figure \ref{fig:Ctflx_Lum} was performed using the AD two-sample test. The result of this test implies that the two distributions are not similar, yet with low significance ($p$-value $\approx 0.004$). The combination of this result and the finding that the low-luminosity sub-sample is non-Gaussian at a high confidence level whereas the high-luminosity distribution is much more consistent with a Gaussian suggests that there is likely a trend between decreasing luminosity and increasing amplitude of \hbox{X-ray} variability. Making additional cuts in luminosity (e.g. cutting equally into thirds) could help to further highlight differences in extreme \hbox{X-ray} variability at different luminosities; however, our sample is not large enough to perform such an analysis and obtain statistically significant conclusions. It would also be instructive to split each timescale sub-sample by luminosity to investigate further this point once a larger sample can be generated. 

We then examined the symmetry of the low- and high-luminosity samples using the AD two-sample test and measured their kurtosis. Both distributions are consistent with being symmetric around zero (see Table \ref{tab:stats}), and the excess kurtosis of the low-luminosity sample ($k = 5.106$) is consistent with a tail-heavy distribution, whereas the high-luminosity sample kurtosis ($k = 0.780$) is only slightly larger than what is expected for a Gaussian distribution given the uncertainty (see Table \ref{tab:stats}). This result again suggests that the low-luminosity sub-sample more frequently displays larger variations than would be expected from Gaussian random fluctuations, and thus the extreme \hbox{X-ray} variations are likely driven by additional physical mechanisms. While the kurtosis of the high-luminosity sub-sample is positive, the value is only marginally larger than what is expected for a Gaussian distribution. An additional physical mechanism might be driving the extreme \hbox{X-ray} variability in this sub-sample; however, the tails of this distribution may also be driven largely by random fluctuations over the timescales probed by this analysis. More data are needed to test further whether or not the extreme \hbox{X-ray} variability in high-luminosity quasars is randomly driven. We find that the spreads in the distributions of the low- and high-luminosity samples are MAD $=0.102$ and MAD $=0.103$, respectively, which suggests that extreme $5\sigma_{\rm MAD}$ \hbox{X-ray} variability occurs at count flux ratios of $\approx 5.71$ and $\approx 5.80$, respectively.

\subsection{Frequency of extreme \hbox{X-ray} variability}\label{sec:binomial}

A primary purpose of this investigation is to quantify the frequency of intrinsic extreme variations in \hbox{X-ray} flux for typical quasars to understand better the nature of the \hbox{X-ray} coronal region and innermost accretion flow. To estimate the frequency of extreme \hbox{X-ray} variation, we adopted the statistical method from Section 3.3.3 of \citet{Gibson2012}. This technique employs binomial statistics to estimate the probability that the magnitude of log$_{10}$(count flux ratio) in a quasar is greater than a given value, $|x|$. If $N$ is the number of measurements in our sample that satisfy this condition, then the probability that a quasar has log$_{10}$(count flux ratio) $> |x|$ can be computed by solving the binomial equation assuming that there is a 95\% binomial probability of finding $N+1$ such objects. The frequencies calculated in this investigation include the intrinsic variations of the quasars as well as the variations due to the measurement errors, which are difficult to model, and thus the frequency of extreme \hbox{X-ray} variability should be considered as upper limits. Insofar as the variations in the measurement uncertainties are small, the upper limits on the fraction of extreme \hbox{X-ray} variability found in this Section can be representative of the true fraction; however we will conservatively report them as upper limits. Moreover, the estimated frequencies presented in this Section are only applicable to quasars that share the same characteristics as our quasar sample (e.g. non-BAL, radio-quiet quasars). 

Ideally this analysis would be performed entirely with well-measured count flux ratios; however, there were a few objects in this work that were not detected in \hbox{X-ray}s, and thus the 90\% confidence limits on the count flux ratio are reported. We incorporated these limits into the calculation of the frequency in two different ways. First, we conservatively assume that all of the limits maximally vary, and are given a value equal to the maximum count flux ratio found among the full population. This method assumes that all of the limits are extremely \hbox{X-ray} variable; however, this is a conservative approach because, in some cases, even a slight decrease in the \hbox{X-ray} flux of a quasar could cause the quasar to be fainter than the limiting flux threshold of a \Chandra observation and thus it can no longer be well detected. A likely more realistic method of incorporating the \hbox{X-ray} limits (hereafter, the median method) is to evaluate the log$_{10}$(count flux ratio) of the limits with respect to the ratios of objects with similar $\Delta t$ values in Figure \ref{fig:Ctrt_all}. For each of the limits depicted in Figure \ref{fig:Ctrt_all}, we compare the count flux ratio (using the 90\% confidence limit as the ``true'' count value) with nearby count flux ratio measurements (i.e,\ $\Delta t$ is within a factor of three). If the count flux ratio of the limit is consistent within $1\sigma$ of its neighbors, we consider the limit to vary as the median of the neighboring objects. If the count flux ratio of the limit exceeds the $1\sigma$ range, we assume maximal \hbox{X-ray} variability as in the conservative method. While this method may not be precise, we find it unlikely that {\emph{all}} of the \hbox{X-ray} non-detections have varied extremely, particularly considering the numerical distribution of limit values in Figure \ref{fig:Ctrt_all}. We report the results of both methods in the following discussion.

\begin{figure*}
\includegraphics[width=0.9\textwidth]{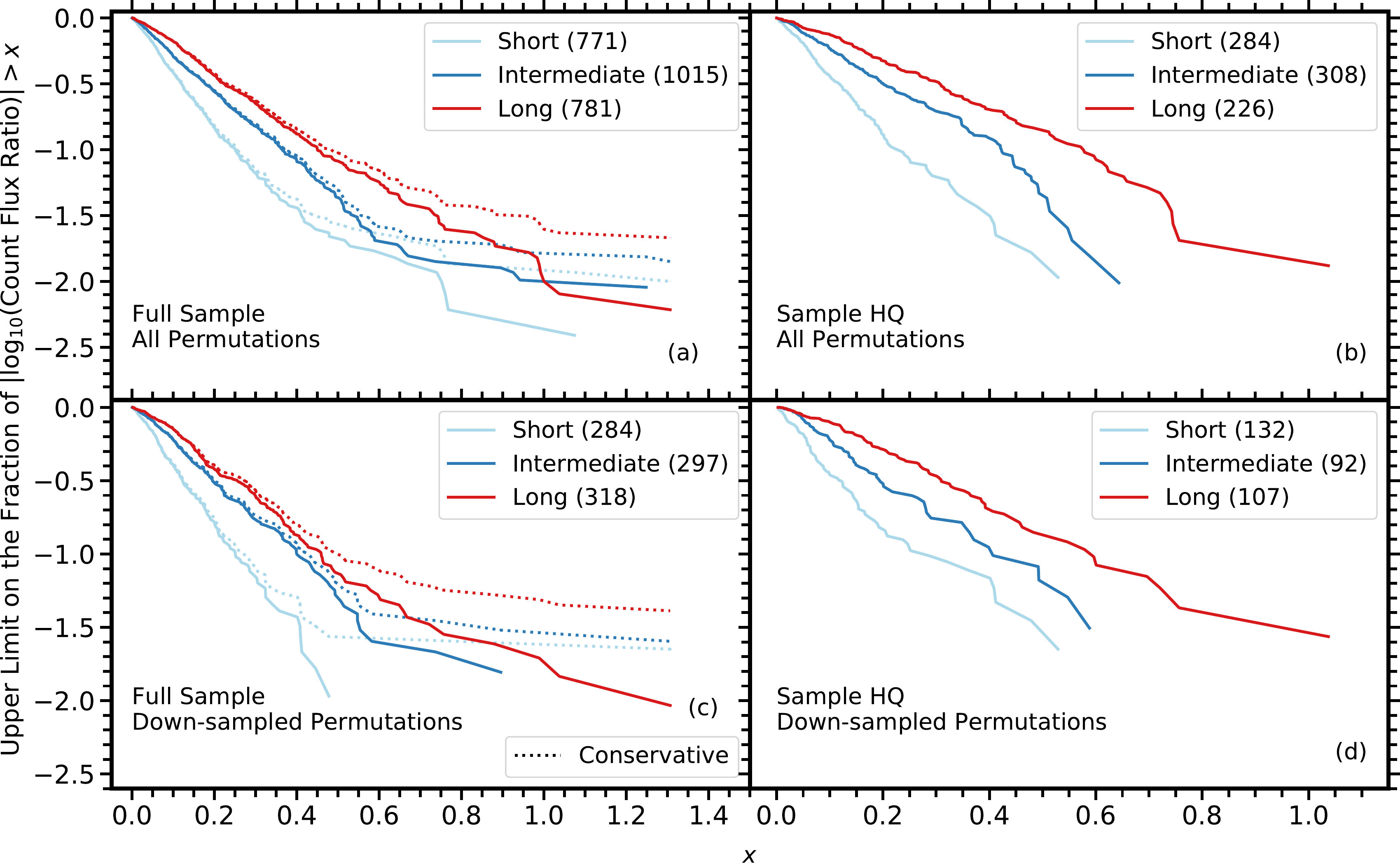}
    \caption{Each panel depicts the upper limit on the frequency at which typical quasars vary by more than a given count flux ratio, $x$, for the four different samples considered in this investigation. Each sample is separated into a short (cyan), intermediate (dark blue), and long (red) timescale sub-sample as defined in Section \ref{sec:dt}. Panels~(a) and (c) also depict the median (solid line) and conservative (dotted line) methods of treating the \hbox{X-ray} limits. The curves in panels~(a) and (b) contain all epoch permutations of each of the quasars in their respective samples. Panels (c) and (d) depict the down-sampled counterparts of the previous two panels and thus are not as heavily weighted toward quasars with many observations. Each of the four panels indicates that the rate of observed extreme intrinsic variation increases with increasing timescale.}
   \label{fig:Binom_dt}
\end{figure*}

We first investigated the upper limit on the frequency of \hbox{X-ray} variation for the quasars in our samples split by timescale from Section~\ref{sec:dt}. Figure \ref{fig:Binom_dt} depicts the results of the binomial analysis on all four of the samples generated in this work. Panels (a) and (b) include all of the epoch permutations for each quasar for the Full sample and Sample HQ, respectively. In panel (a), we also depict the two methods of incorporating the \hbox{X-ray} non-detections into the calculation (solid line, median method; dotted line, conservative method). For clarity, we report the logarithm of the frequency for each given sub-sample. For example, panel~(a) of Figure \ref{fig:Binom_dt} reports that an \hbox{X-ray} variation by a factor of ten occurs for $<10^{-1.6}\approx 2.5\%$ of observations at long timescales using the conservative limit approach and $<10^{-2.0} \approx 1.0\%$ of observations using the median limit approach. From panel (b), which contains no \hbox{X-ray} limits, the same amplitude of variation occurs in \hbox{$<10^{-1.85} \approx 1.6\%$} of observations at long timescales. A comparable estimate of the frequency for this factor of ten variation is $\approx 4.0\%$ using the estimates in \citet{Gibson2012}, and thus our sample provides significantly improved constraints on extreme \hbox{X-ray} variability.

Depicted in panels (c) and (d) of Figure \ref{fig:Binom_dt} are the results of the analysis when performed on the down-sampled versions of the Full sample and Sample HQ, respectively. In Section \ref{sec:dt}, we have defined ``extreme'' variations as the $5\sigma_{\rm MAD}$ deviation from the median for each of the three timescale sub-samples of the Full sample with reduced permutations. From panel (c) of Figure~\ref{fig:Binom_dt}, the upper limit on the frequency of these extreme variations is between $<1.0\%$--$<3.1\%$, $<1.9\%$--$<3.3\%$, and $<1.9\%$--$<4.7\%$ at short, intermediate, and long timescales (where we report the median--conservative method of incorporating the limits).\footnote{The short and intermediate timescale bins do not extend to the $5\sigma$ value when treating the limits using the median method, so we instead report the smallest value of the frequency for these sub-samples.} Panel (d), which is not affected by \hbox{X-ray} limits, reports a frequency between the median and conservative approaches in panel (c) for the long timescale sub-sample ($<3.2\%$); however, the short and intermediate timescale sub-samples do not extend to the $5\sigma$ value computed in Section \ref{sec:lum}, and thus no precise limit can be determined beyond what is found in panel (c). All of the panels in Figure~\ref{fig:Binom_dt} demonstrate that extreme intrinsic \hbox{X-ray} variability occurs more frequently at long timescales than at short timescales, and that it is a very rare phenomenon. Although the frequencies of extreme \hbox{X-ray} variability derived above are small, we do not consider them to be consistent with zero since the kurtosis values of the three timescale distributions, which can also be used as a measure of the frequency of extreme variations, are not consistent with zero (see Section~\ref{sec:dt}). Radio-quiet, non-BAL quasars that are found to exhibit such extreme \hbox{X-ray} variability should therefore be closely monitored to understand better the driving mechanisms of this quantifiably remarkable phenomenon. 
 
\begin{figure*}
\includegraphics[width=0.9\textwidth]{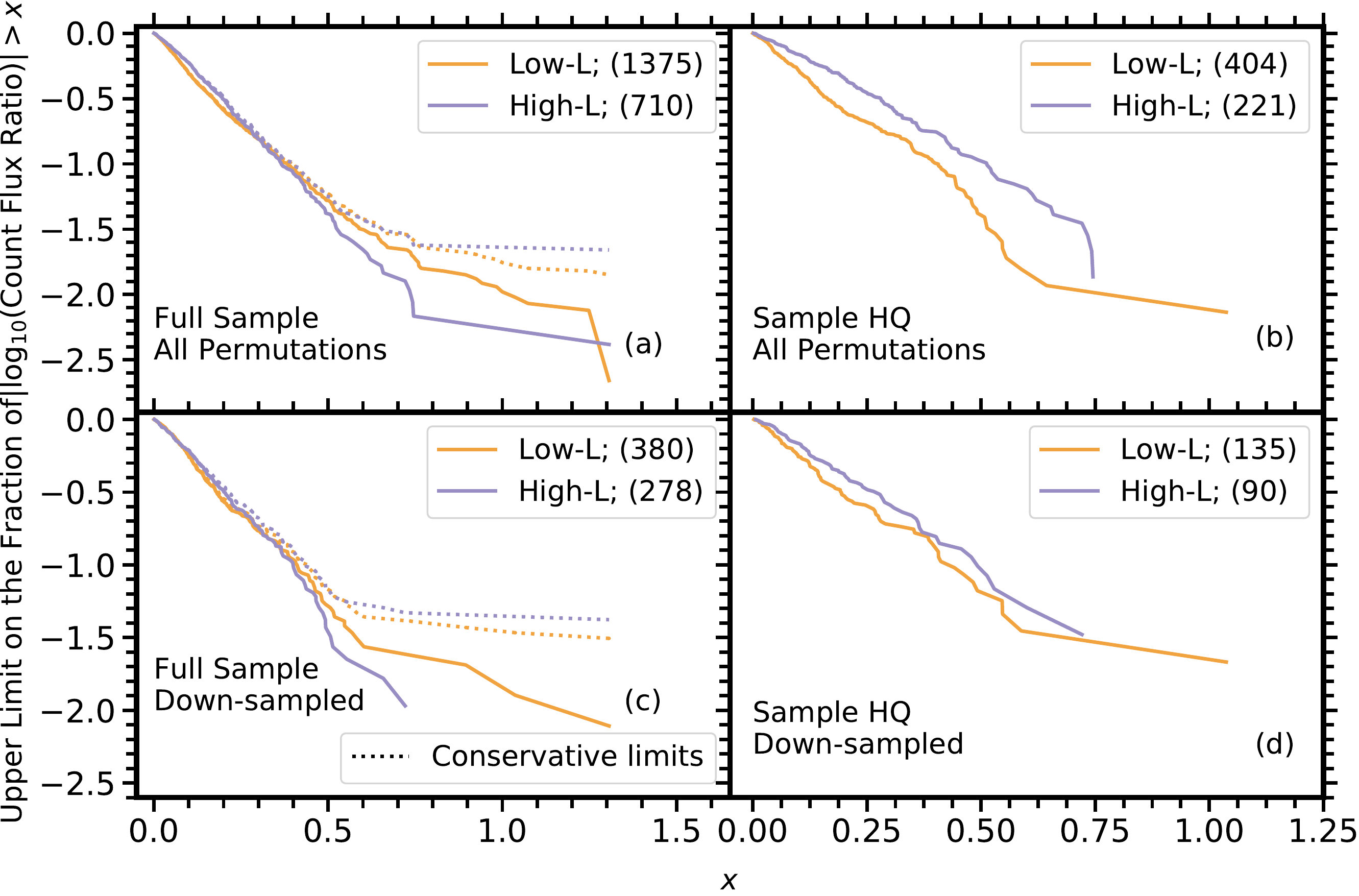}
    \caption{Similar to Figure \ref{fig:Binom_dt}, however instead of splitting by timescale, the samples were split at the median luminosity into high-luminosity (purple) and low-luminosity (orange) sub-samples (see Section \ref{sec:lum}). The upper limits on the frequency in these two sub-samples are much more consistent than when splitting by timescale; however, at large-amplitude variations, the frequency of \hbox{X-ray} variability in the low-luminosity sub-sample is generally larger than that of the high-luminosity sub-sample on similar timescales. The sudden decrease in the frequency at large count flux ratio in the low-luminosity sub-sample depicted in panel (a) is the result of removing some of the extreme variations at long timescales when matching the low- and high-luminosity sub-samples in timescale. }
   \label{fig:Binom_lum}
\end{figure*}

The frequency of extreme \hbox{X-ray} variation in the two luminosity sub-samples presented in Section~\ref{sec:lum} was also calculated. Figure~\ref{fig:Binom_lum} depicts the four samples of quasars in the same manner as in Figure~\ref{fig:Binom_dt}. In all of the panels, the upper limit on the frequency of variation in the low- and high-luminosity sub-samples is very similar for small-amplitude \hbox{X-ray} variations, but as the amplitude increases the frequency of occurrence for the low-luminosity quasars tends to be larger than that of the high-luminosity quasars when the upper limits are incorporated using the median method.\footnote{This is apparently not the case in panel (b); however, we found that a single object with numerous observations is causing the large difference between the two samples. Removing the permutations in panel (d) reduces the effect of this issue.} From panel (c) of Figure~\ref{fig:Binom_lum}, the extreme variations defined in Section \ref{sec:lum} in the high- and low-luminosity sub-samples occur at a rate of $<1.0\%$--$<4.9\%$ and $<2.4$--$<4.0\%$, respectively, where the median--conservative limit approaches are reported as before. Once again, these curves were derived from the luminosity sub-samples that were designed to have similar timescale distributions. If the physical driver of extreme variations is luminosity-dependent, then the high-luminosity quasars might vary extremely ($\gtrsim 10\times$) at a similar frequency as the low-luminosity quasars, but on a timescale even longer than that probed by this analysis. Splitting the luminosity bins by the median timescale ($\Delta t = 6.76 \times 10^{6}$ s), we find that the low-luminosity, long timescale sub-sample has the largest frequency of extreme \hbox{X-ray} variability ($< 3.0 \%$ using the median method); however, additional data are required to increase the sizes of the samples to confirm this result.

Figures \ref{fig:Binom_dt} and \ref{fig:Binom_lum} both demonstrate that extreme \hbox{X-ray} variability occurs very infrequently in the sample that was constructed in this work. The low rate of occurrence of these events likely indicates that extremely \hbox{X-ray} variable quasars are rare objects in the overall majority radio-quiet quasar population. Another possibility, at least in principle, is that the duty cycle of extreme \hbox{X-ray} variability is low and the \hbox{X-ray} light-curves are not sampled sufficiently densely to capture these events when they occur. Our investigation, however, was designed to include a large sample of \hbox{X-ray} observations of quasars that span a wide range of rest-frame timescales (up to $\approx 12$ years) in order to detect extreme \hbox{X-ray} variability even in the case of a small duty cycle. Unless the duty cycle is very small, it is likely that the low frequencies computed in this Section reflect the fact that extremely \hbox{X-ray} variable quasars are rare objects in the overall quasar population.

In Section \ref{sec:EXV_objects}, we report some \hbox{X-ray} and optical/UV emission-line properties of three SDSS quasars that we discovered in our work to have exhibited extreme \hbox{X-ray} variability according to the condition derived in Section \ref{sec:binomial} for long timescales (a flux ratio $>9.85$). Each of these quasars was found to have notably hard \hbox{X-ray} spectra in at least one epoch which generally indicates that the \hbox{X-ray} spectrum is absorbed (though, the quasar J1420+5254 may also show evidence of intrinsic extreme \hbox{X-ray} variability; see Section \ref{sec:J1420}). \hbox{X-ray} absorption in quasars can be complex (e.g. \citealt{Gallagher2002, Gallagher2006}), and can come from many different physical sources. For example, such absorption could possibly be attributed to a BAL wind (and any associated shielding gas) since blue, type 1 quasars with heavy \hbox{X-ray} absorption are typically found to possess BALs (e.g. \citealt{Brandt2000}). If this is the case, the extreme \hbox{X-ray} variability for these quasars might be the result of a larger scale environment change and not an intrinsic change in the corona and innermost accretion flow, and thus following Section \ref{sec:intro} these quasars should be removed from our analysis. Alternatively, the absorption in these quasars could potentially be due to shielding by a thick inner accretion disk. At large viewing angles, the central \hbox{X-ray} corona may be absorbed by the thick disk, but changes in the covering factor of the corona by the disk due to variations in the size of the corona or azimuthal asymmetries of the inner disk (e.g. \citealt{Liu2019, Ni2020}) may cause an extreme \hbox{X-ray} variability event. In this case, the extreme variations of these quasars would be due to changes in the innermost accretion flow, and thus we should retain these quasars in our analysis. 

\begin{figure}
\includegraphics[width=\columnwidth]{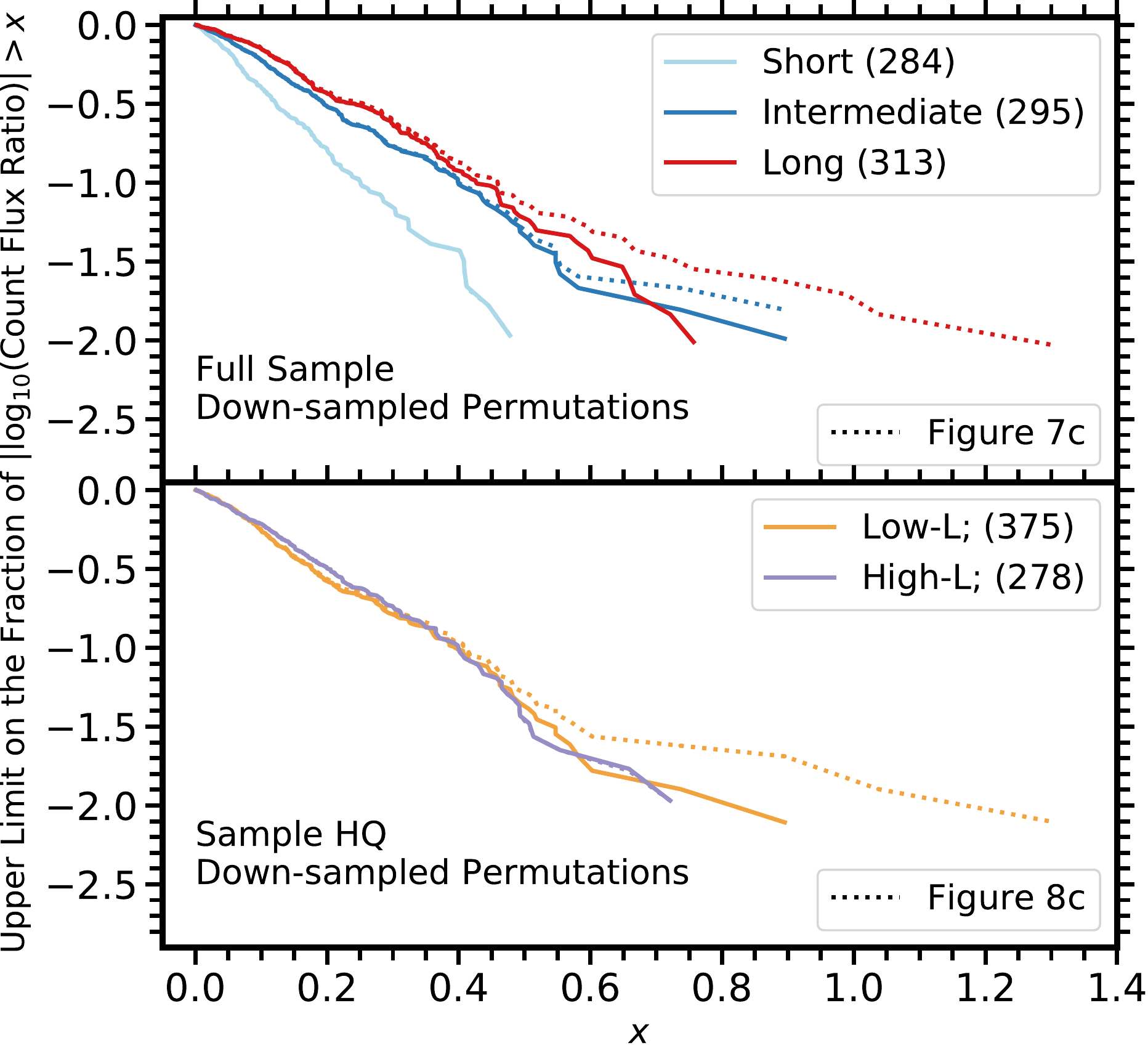}
    \caption{The frequencies of \hbox{X-ray} variability in the three timescale bins from Section \ref{sec:dt} (top panel) and two luminosity bins from Section \ref{sec:lum} (bottom panel) after removing the three quasars that we conservatively assume to extremely vary due to \hbox{X-ray} absorption by a BAL wind (see Section \ref{sec:EXV_objects}). The solid curves depict the frequencies after removing these quasars and the dotted curves depict the results presented in panel (c) of Figure \ref{fig:Binom_dt} for the timescale sub-samples (top panel) and Figure \ref{fig:Binom_lum} for the luminosity sub-samples (bottom panel). \hbox{X-ray} limits are treated using the median method for each curve. After removing these potentially \hbox{X-ray} absorbed quasars, we find that the frequency of extreme \hbox{X-ray} variability marginally decreases in the low-luminosity, intermediate-timescale, and long-timescale sub-samples. The frequencies depicted in both panels are only mildly smaller than those in Figures \ref{fig:Binom_dt} and \ref{fig:Binom_lum} which illustrates the robustness of the results of our full analysis and further demonstrates that extreme \hbox{X-ray} variability in radio-quiet, non-BAL quasars is a rare phenomenon.}
   \label{fig:Binom_rmabs}
\end{figure}

We cannot directly confirm either scenario for our three newly discovered extremely \hbox{X-ray} variable quasars without follow-up spectroscopic observations of their rest-frame UV emission (note these quasars lie at $z<1.7$, so the SDSS spectra do not cover the \ion{C}{iv} region). We therefore conservatively assume in this paragraph that all three quasars that may be \hbox{X-ray} absorbed in at least one epoch are BAL quasars and thus they should be removed from the sample following Section~\ref{sec:RL_BAL}. Figure \ref{fig:Binom_rmabs} depicts the frequency of extreme \hbox{X-ray} variability after removing these objects. The top panel of Figure~\ref{fig:Binom_rmabs} depicts the three timescale bins in the reduced epoch permutation sample akin to panel (c) of Figure \ref{fig:Binom_dt}, and the bottom panel similarly presents the two luminosity bins as shown in panel (c) of Figure \ref{fig:Binom_lum}. In both panels, we treat the \hbox{X-ray} limits using the median method. Figure \ref{fig:Binom_rmabs} clearly illustrates that the frequencies of the intermediate-timescale, long-timescale and the low-luminosity bin decrease somewhat compared to those in Figures \ref{fig:Binom_dt} and \ref{fig:Binom_lum}, respectively, whereas the frequencies computed for the short-timescale and high-luminosity sub-samples are not affected by the removal of these quasars. The mild decrease illustrates the robustness of the results in Figures~\ref{fig:Binom_dt} and \ref{fig:Binom_lum} and further confirms that intrinsic extreme \hbox{X-ray} variability is a rare phenomenon. 

Although the frequency of intrinsic extreme \hbox{X-ray} variations is very low, our analysis indicates that they are not likely an artifact of statistical fluctuations, rather seeming to be driven by at least one additional physical mechanism. Models of the corona and innermost accretion flow must therefore reproduce the rarity of these extreme \hbox{X-ray} variability events without considering them to be merely random fluctuations of the \hbox{X-ray} flux.


\section{Extremely \hbox{X-ray} variable objects}\label{sec:EXV_objects}
During the course of this investigation, we discovered three quasars that have varied extremely (by $\gtrsim 9.85\times$) in their full-band count flux consistent with the threshold derived in Section~\ref{sec:binomial}. These three quasars were used in the analyses reported in Section~\ref{sec:binomial} (aside from Figure \ref{fig:Binom_rmabs}, where they were explicitly removed). All three of these quasars have 2500 \AA\ luminosity (derived from the single-epoch $i$-band magnitude in the SDSS DR14Q catalog) consistent with our low-luminosity bin. Here we present a brief overview of the basic properties of these three objects. More detailed investigations into the relationship between the UV/optical and \hbox{X-ray} light-curves of these objects will be reserved for future work.

\subsection{SDSS J141531.93+112850.7}\label{sec:J1415}
There were only two observations of SDSS J1415+1128 ($z = 0.360$; log$_{10}(L_{2500}) = 29.11$ erg s$^{-1}$) in the \Chandra archive that satisfied our requirements, with the first taken in April 2000 ($\approx 18$ ks exposure) and the second in March 2005 ($\approx 84$ ks exposure). A third observation was found from April 2013; however, the quasar coincided with an ACIS chip-gap, and thus the observation was automatically removed from our analysis. Between 2000 and 2005 ($\Delta t$ $\approx 1.15 \times10^{8}$~s; $\approx 1330$ days), this object increased in full-band count flux by a factor of $\approx 11$. The measurement from the observation in 2013 suggests that the flux then decreased by a factor of $\approx 2.7$; however, a more detailed analysis is required since in 2013 the quasar lies on a chip gap and thus the measurement is likely highly uncertain. 

Provided that the UV luminosity does not change significantly over time, the expected {\hbox{X-ray}-to-optical} spectral slope is $\alpha_{\rm ox} = -1.37$, using the relationship between log$_{10}(L_{2500})$ and $\alpha_{\rm ox}$ from \citet{Just2007}.\footnote{$\alpha_{\rm ox} = 0.3838\times {\rm{log}}_{10}(f_{2\ \rm keV}/f_{2500})$, where $f_{2\ \rm keV}$ and $f_{2500}$ are the flux densities at 2 keV and 2500 \AA, respectively.} The measured values from the 2 keV flux density are $\alpha_{\rm ox} = -1.87$  and $\alpha_{\rm ox} = -1.46$ for the first and second epoch, respectively. Comparing the measured values of $\alpha_{\rm ox}$ with the expected value indicates that the \hbox{X-ray} emission from the quasar was significantly weaker than expected in the first epoch ($\Delta\alpha_{\rm ox} = -0.50$) but consistent with being \hbox{X-ray} normal in the second epoch ($\Delta\alpha_{\rm ox} = -0.09$) considering the intrinsic dispersion of $\alpha_{\rm ox}$ values in Table 5 of \citet{Steffen2006}. Using the hard-to-soft band ratios and the \Chandra Portable, Interactive Multi-Mission Simulator (PIMMS)\footnote{\url{https://cxc.harvard.edu/toolkit/pimms.jsp}} software, we measured the effective power-law photon index, $\Gamma$, for the faint and bright epochs (assuming only Galactic absorption). The first, fainter epoch has a flat effective photon index ($\Gamma = 1.0^{+0.5}_{-0.5}$), albeit with a large uncertainty, whereas the second, bright epoch has a steeper effective photon index ($\Gamma = 1.85^{+0.09}_{-0.10}$) akin to that of a typical quasar ($\Gamma = 1.8$; e.g. \citealt{Scott2011}). The hard spectrum in the first epoch suggests the presence of \hbox{X-ray} absorption which may indicate that a BAL wind is present in this quasar that we are unable to observe with SDSS. The quasar gets brighter and softer in the second epoch as expected if a simple \hbox{X-ray} absorber moves out of the line-of-sight to the \hbox{X-ray} continuum. If a BAL wind is present, the extreme \hbox{X-ray} variability would be the result of a larger scale environment change and not an intrinsic change in the corona and innermost accretion flow; therefore, we remove this quasar when performing the statistical analysis presented in Figure~\ref{fig:Binom_rmabs}.  

The \ion{H}{$\beta$} emission-line width in this blue, type 1 quasar is not narrow (FWHM $\approx 6000\ \rm{km\ s^{-1}}$) and the [\ion{O}{iii}] 5007\AA\ emission-line strength is consistent with that of a typical quasar (EW $\approx 19$~\AA) according to our fits to the SDSS spectrum. This object is also reported in the 2XMM serendipitous source catalog \citep{Watson2009} from {\emph{XMM-Newton}} observations in 2001 and 2002 which are between the \Chandra observations. Both {{\emph{XMM-Newton}} observations confirm that the source continued getting brighter between the \Chandra measurements. New observations of this object are required to determine if the flux has continued to increase, or if the flux peak occurred in 2005.

\subsection{SDSS J141751.13+522311.0}\label{sec:J1417}
There are five serendipitous \Chandra observations of SDSS J1417+5223 ($z = 0.281$; log$_{10}(L_{2500}) = 28.76$ erg s$^{-1}$) with three in August/September 2002 ($\approx 122$ ks of stacked exposure time) and two more in September/October 2014 ($\approx 34$ ks of stacked exposure time) as depicted in Figure \ref{fig:1417_LC}. The \hbox{X-ray} flux of this quasar has increased between these two epochs ($\Delta t$ $\approx 3.0 \times10^{8}$ s; $\approx 3472$ days) by a factor of $\approx 9.9$. From the single-epoch UV luminosity and the relationship in \citet{Just2007}, the expected \hbox{X-ray}-to-optical spectral slope is $\alpha_{\rm ox} = -1.32$ for this quasar. In 2002, this quasar was \hbox{X-ray} weaker than expected, with a minimum $\alpha_{\rm ox} = -1.80$ ($\Delta\alpha_{\rm ox} = -0.48$); however, in 2014 the \hbox{X-ray} flux significantly increased, corresponding to a value of $\alpha_{\rm ox} = -1.43$ ($\Delta\alpha_{\rm ox} = -0.09$) which indicates this quasar became \hbox{X-ray} normal. We measured the effective photon indices using the stacked observations in 2002 and 2014, respectively, since there is only a small temporal separation between the measurements in these two epochs. We find a similar effective photon index between 2002 ($\Gamma = 1.23^{+0.17}_{-0.20}$) and 2014 ($\Gamma = 1.13 ^{+0.09}_{-0.09}$). The consistently hard spectrum again suggests that the \hbox{X-ray}s are being absorbed in both epochs, though it is puzzling that the source has retained a hard \hbox{X-ray} spectrum when its flux returns to a nominal level for a typical quasar. As before, if a BAL wind is the primary reason for the absorption, this quasar must be removed from the analysis in Section \ref{sec:binomial}, which we do when performing the statistical analysis for Figure~\ref{fig:Binom_rmabs}. 

Our fit of the SDSS spectrum again indicates that this quasar has a broad \ion{H}{$\beta$} emission line (FWHM $\approx 15000\ \rm{km\ s^{-1}}$), and it exhibits similar [\ion{O}{iii}] 5007\AA\ emission strength compared with that of a typical quasar (EW $\approx 11$ \AA). This object is also reported in the 3XMM serendipitous catalog \citep{Rosen2016} and was included in the the AGN \hbox{X-ray} variability work of \citet{Vagnetti2016}. The timescale between the {\emph{XMM-Newton}} observations from July 2000 that were used in that investigation is only $\approx 1.3$ days, and no noticeable difference in the \hbox{X-ray} flux was reported. The full-band flux value in the {\emph{XMM-Newton}} observation in 2000, however, is {\mbox{$\approx 3\times$}} fainter than the 2002 \Chandra observation which suggests that this quasar has increased in flux by a total factor of $\approx 29$ from 2000 to 2014. One additional  {\emph{XMM-Newton}} observation is reported in the 3XMM catalog that was not included in the \citet{Vagnetti2016} investigation from 2014 which has a full-band \hbox{X-ray} flux value consistent with the 2014 \Chandra observation. The three \Chandra observations in 2002 and the two in 2014 also suggest that this quasar does not vary significantly on short timescales, and thus this appears to be the first reported instance of extreme \hbox{X-ray} variability from this quasar.

\begin{figure}
\includegraphics[width=\columnwidth]{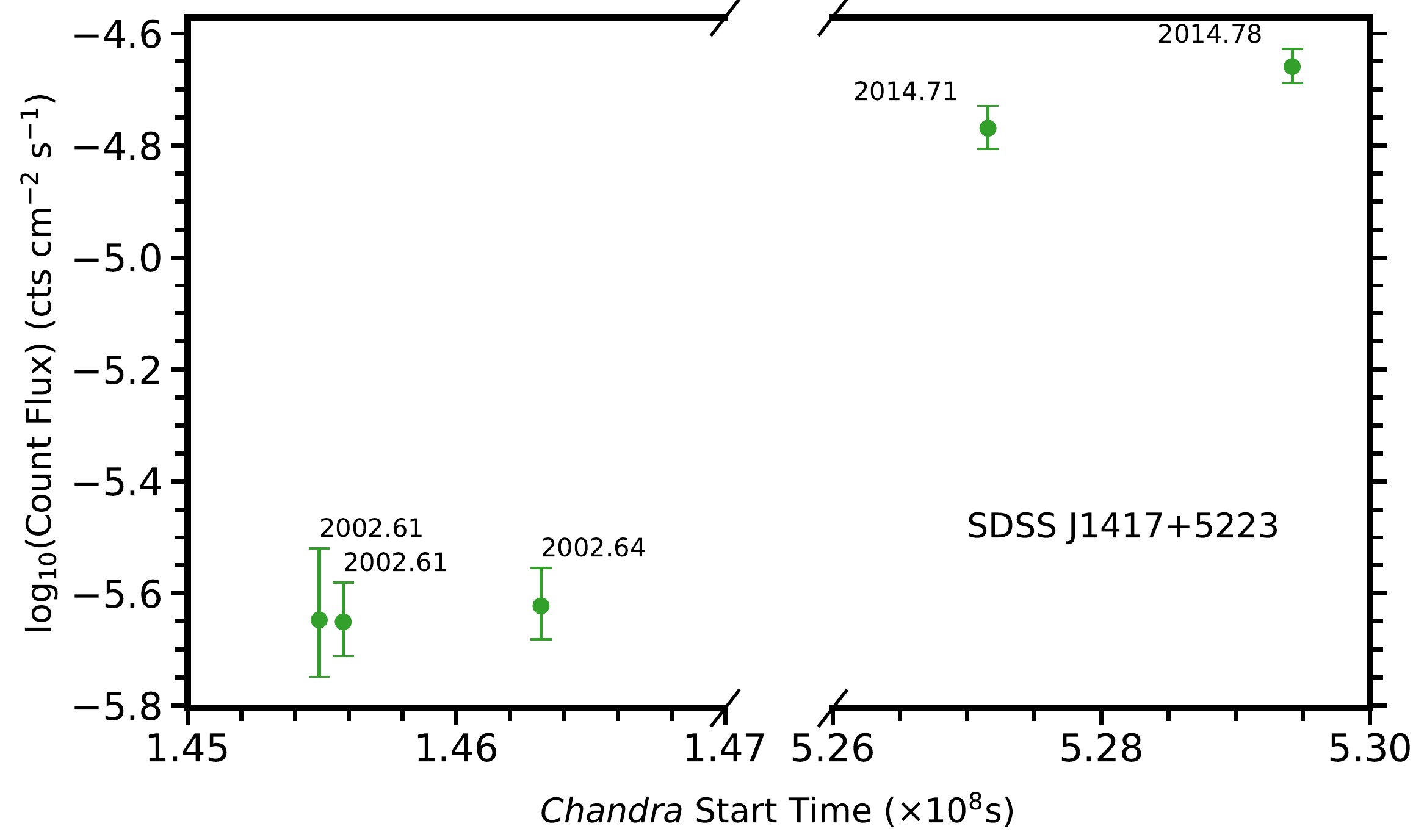}
    \caption{The full-band count flux as a function of the \Chandra start time for the five observation epochs of SDSS J1417+5223. The values above the data points report the decimal year of the \Chandra observation. The maximum count flux increased by a factor of $\approx 9.8$ over $\approx 3472$ rest-frame days for this quasar.}
   \label{fig:1417_LC}
\end{figure}

\subsection{SDSS J142037.84+525452.8}\label{sec:J1420}

SDSS J1420+5254 ($z = 1.269$; log$_{10}(L_{2500}) = 30.05$ erg s$^{-1}$) was observed 22 times with \Chandra as part of the \hbox{AEGIS-X} survey (e.g.\ \citealt{Laird2009}), which was a deep observation of the Extended Groth Strip between 2005 and 2008. The full-band \hbox{X-ray} light-curve for this object is depicted in Figure \ref{fig:1420_LC}, where the full-band count flux values of the \hbox{X-ray} detections (upper limits) are depicted by the green circular points (black downward arrows). We found that the largest factor by which the full-band count flux of this object varied was $\approx 21$ on a timescale of {\mbox{$\Delta t$ $\approx 3.8 \times10^{7}$ s}} (439 days) in the rest frame. This object also exhibited short-timescale extreme variability by a factor of $\approx 12$ on a timescale of {\mbox{$\Delta t$ $\approx 2.0 \times10^{5}$ s}} (2.73 days) in the rest frame. This quasar was not detected by \Chandra in three observations; therefore, the X-ray flux might have decreased by an even larger factor (see Figure \ref{fig:1420_LC}). 

Generally, the number of full-band counts in each epoch is low, which yields multiple instances in which this quasar is not detected in either the soft or the hard band, but is detected in the full band, making it difficult to estimate the shape of the \hbox{X-ray} spectrum. For example, the epoch with the lowest count flux value, which occurred in March 2005, did not have sufficient counts in the soft band to be detected, but was detected in the hard and full bands. We derived a limit on the effective photon index to be $\Gamma < 0.5$ and, using the full-band counts, we found that $\alpha_{\rm ox} = -1.90$ in this epoch, which suggests that it is X-ray weak compared to the expectation ($\alpha_{\rm ox} = -1.50$). In the bright state (December 2007), we found that $\alpha_{\rm ox} = -1.34$ ($\Delta\alpha_{\rm ox} = 0.16$) and we found a steep effective photon index ($\Gamma = 2.5^{+0.4}_{-0.5}$) consistent with that of a typical quasar. As before, the transition from the flat, X-ray weak state to the steep, X-ray brighter state may suggest that this quasar is absorbed.

Since this quasar was observed over many more epochs that the previous objects, we further investigated the X-ray variability of this quasar by limiting our consideration to the epochs in which the quasar was detected in all three bands (seven epochs). After limiting to only these epochs, the quasar still exhibits extreme X-ray variability, dropping from the aforementioned high state by factor of $\approx 14$ on a timescale of {\mbox{$\Delta t$ $\approx 7.1 \times10^{6}$ s}} (82 days) in the rest frame. We found that $\alpha_{\rm ox} = -1.78$ in this well-detected low state (June 2008) and that the effective photon index remained steep ($\Gamma = 1.9^{+0.9}_{-1.1}$) as in the bright state. Unlike before, this extreme decrease in \hbox{X-ray} flux but similarly steep effective photon index may suggest that this variability may be intrinsic to the coronal region and innermost accretion flow. It therefore seems possible that the extreme X-ray variations are caused by either absorption, intrinsic fluctuations, or a combination of the two. Neither the rest-frame optical nor the \ion{C}{iv} emission-line region are present in the SDSS spectrum; thus, we cannot determine if this quasar exhibits similar emission-line properties to the other extremely X-ray variable quasars (e.g.\ \citealt{Liu2019}), nor can we definitively determine if there is a BAL present in the spectrum. Although there may be evidence of intrinsic X-ray variability, this quasar exhibited a flat \hbox{X-ray} spectrum in at least one epoch; therefore, we conservatively assumed that the variations were caused by BAL absorption, and thus removed this object in Figure \ref{fig:Binom_rmabs}. Additional spectral observations of both the rest-frame optical and the \ion{C}{iv} emission-line region are required to understand better the nature of the \hbox{X-ray} variability of this object. 

\begin{figure}
\includegraphics[width=\columnwidth]{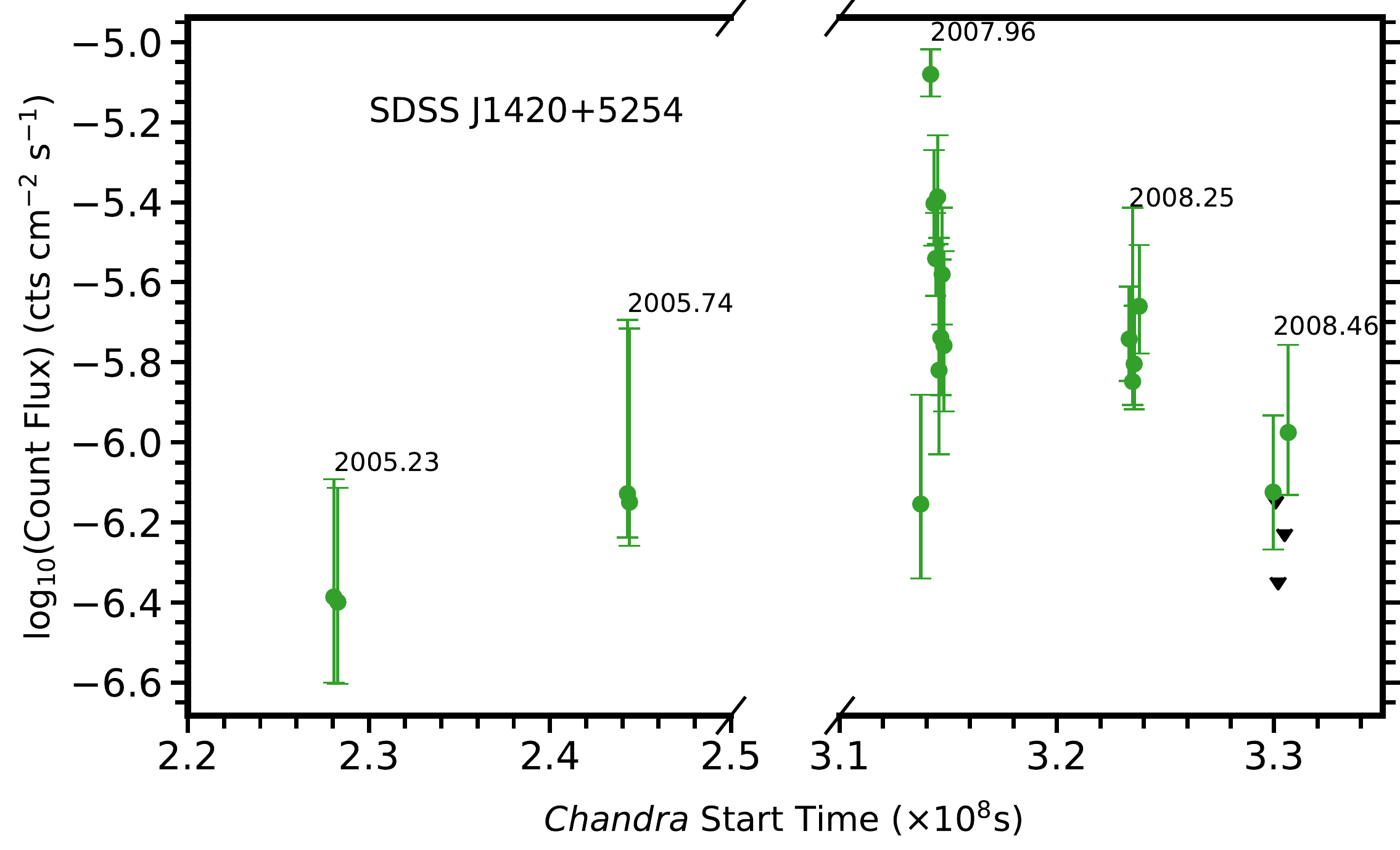}
    \caption{The full-band count flux as a function of the \Chandra start time for the 22 observation epochs of SDSS J1420+5254. The values above the data points report the approximate decimal year of the serendipitous \Chandra observations. The count flux increased maximally by a factor of $\approx 21$ over $\approx 439$ rest-frame days for this quasar. }
   \label{fig:1420_LC}
\end{figure}


\section{Summary and Future Work}
In this investigation, we compiled a large sample of quasars that had multiple, serendipitous \Chandra observations in order to constrain the frequency of extreme \hbox{X-ray} variability intrinsic to the corona and innermost accretion flow among typical quasars. Radio-loud and BAL quasars were removed from the sample since variations of their \hbox{X-ray} flux can arise from larger scale environmental effects (e.g.\ a jet contribution or wind-associated absorption). We generated a Full sample of 1598 sensitive observations of 462 quasars, along with a high-quality sample of 583 observations of 185 quasars, in which all observations resulted in an \hbox{X-ray} detection. The light-curves of frequently observed quasars in these data sets were also down-sampled to reduce the weight of any one quasar with many observations in the sample. The Full sample of quasars with reduced epoch permutations was split into three timescale bins (short, intermediate, and long) and two luminosity bins (faint and bright) to investigate the frequency of extreme \hbox{X-ray} variations as a function of these properties. The main results of this analysis are the following: 

\begin{enumerate}[i]

\item All three timescale bins were not consistent with a Gaussian distribution according to an AD test of normality, which suggests that the extreme \hbox{X-ray} variations are not a result of random fluctuations in the coronal emission. Rather, it is likely that an additional physical mechanism acting occasionally drives this extreme variability. Additionally the long timescale bin was found to exhibit larger amplitude \hbox{X-ray} variations more frequently than the short or intermediate timescale bins compared to what would be produced by random fluctuations (Section \ref{sec:dt}).

\item We used the median-absolute-deviation to define ``extreme" $5\sigma_{\rm MAD}$ variations in our three timescale sub-samples. Our analysis indicates that extreme \hbox{X-ray} variations occur when the flux changes by a factor of $\approx 3.19, \approx 6.54, \rm{and} \approx 9.85$ at short ($\Delta t \lesssim 5.3$ days), intermediate ($5.3 \lesssim \Delta t \lesssim 294$ days), and long ($\Delta t \gtrsim 294$ days) timescales, respectively (Section \ref{sec:dt}). At long timescales, we confirm that the factor of ten used in the literature to define extreme \hbox{X-ray} variations is a reasonable choice for the majority population of quasars.

\item The distribution of log$_{10}$(count flux ratio) in the low-luminosity sub-sample is not consistent with a Gaussian at a high confidence level, whereas the high-luminosity sub-sample is much more consistent with a Gaussian according to an AD normality test. This again suggests that the extreme \hbox{X-ray} variations in low-luminosity quasars arise from an additional physical mechanism whereas the high-luminosity sample is either more consistent with random fluctuations, or the physical mechanism operates on longer timescales than at lower-luminosity (Section \ref{sec:lum}). The extreme $5\sigma_{\rm MAD}$ \hbox{X-ray} variations occur when the count flux ratio is $\approx 5.71$ and $\approx 5.80$ in the low-luminosity and high-luminosity sub-samples, respectively.

\item Using binomial statistics, we estimated the frequency of extreme \hbox{X-ray} variations for the three timescale bins (Figure \ref{fig:Binom_dt}) and the two luminosity bins (Figure \ref{fig:Binom_lum}). The frequency of extreme \hbox{X-ray} variations clearly increases with timescale, and it appears to be larger in low-luminosity quasars (Section \ref{sec:binomial}).

\item The results of the binomial analysis indicate that the frequency of extreme \hbox{X-ray} variations in the short, intermediate, and long timescale bins are {\mbox{$<1.0\%$, $<1.9\%$, and $<1.9\%$}}, respectively, using the definition of extreme variability at the $5\sigma_{\rm MAD}$ deviation and using the median method of treating the \hbox{X-ray} limits. Furthermore, extreme \hbox{X-ray} variations in the low-luminosity and high-luminosity sub-samples occur in $< 2.4\%$ and $<1.0\%$  of observations (Section~\ref{sec:binomial}). Extreme \hbox{X-ray} variability that is intrinsic to the coronal region is thus a rare phenomenon.

\item Finally, we found three quasars that have exhibited extreme \hbox{X-ray} variations on long timescales. All three of the quasars have luminosities consistent with our low-luminosity bin and have varied by more than a factor of $9.85$. These quasars might be \hbox{X-ray} absorbed in at least one epoch (see Section \ref{sec:EXV_objects}), and that absorption may be from an undetectable BAL wind. Conservatively assuming that these three are BAL quasars, we re-computed the frequencies of extreme \hbox{X-ray} variability. Figure \ref{fig:Binom_rmabs} depicts that the frequencies of extreme \hbox{X-ray} variability after removing the three potential BAL quasars marginally decreases, further confirming the rarity of these extreme events and the robustness of our statistical conclusions. Additional observations of these objects are needed to probe better their variability properties.

\end{enumerate}

This work could be extended in a few different ways. One way to increase the sample size would be to relax the $i$-band magnitude constraint imposed; however, doing so would significantly increase the number of \hbox{X-ray} count limits in the sample and thus more advanced statistical analyses would be required. Such analyses could not be performed in this work since our sample contained upper and lower limits. Future investigations will have to create a new metric for measuring the variability amplitude such that the limits point in a uniform direction; however, careful consideration must me made to ensure that the new measurement does not introduce biases. Another extension of this work that can be done in the near future will be to incorporate data from the eight epochs of the {\emph{eROSITA}} all-sky survey \citep{Merloni2012} which will greatly enhance the sample size of bright quasars that can be used to increase the statistical power of the results. A brighter $i$-band magnitude limit will need to be imposed since the median flux of the quasars in our investigation is $\approx 3\times$ fainter than the projected limiting flux values of a single {\emph{eROSITA}} epoch; however, the large area of the survey should add many more quasars to the sample. The additional data will be particularly useful for understanding better the difference in the frequency of extreme \hbox{X-ray} variability between the low-luminosity and high-luminosity quasars. 

Another follow-up investigation would be to compare the \hbox{X-ray} variability properties, including the frequency at which extreme \hbox{X-ray} variability occurs, of radio-loud and BAL quasars to our sample of typical radio-quiet quasars. A recent \hbox{X-ray} investigation of radio-loud quasars has suggested that the \hbox{X-ray} emission in some populations of radio-loud quasars still largely originates in the corona instead of being enhanced by the jet, as has been generally accepted \citep{Zhu2020}. Comparing the \hbox{X-ray} variability properties, including the rates of extreme \hbox{X-ray} variability, of radio-loud quasars and our typical radio-quiet quasars, could provide further insights into the origin and behavior of the \hbox{X-ray} emission in these radio-loud quasars. 

Additionally, a more detailed investigation of the three extremely \hbox{X-ray} variable quasars discussed in Section \ref{sec:EXV_objects} would aid our understanding of the physical nature of these extreme objects. A joint investigation of the extreme \hbox{X-ray} variability along with the variability (or lack thereof) at other wavelengths could provide further insight into the connections between the \hbox{X-ray} and optical/UV emission in quasars (e.g. \citealt{Liu2019, Ni2020}). Since none of these quasars were observed recently in \hbox{X-rays}, obtaining new \Chandra (or other \hbox{X-ray}) observations of these quasars would also be useful to determine how their \hbox{X-ray} fluxes have changed since their final observations. Obtaining rest-frame UV spectra would also allow us to determine if they are BAL quasars and measure their \ion{C}{iv} emission lines.

\section*{Acknowledgements}

We thank the referee for a constructive and prompt report. JDT, WNB, SZ, and QN acknowledge support from NASA ADP grant 80NSSC18K0878, \Chandra \hbox{X-ray} Center grant G08-19076X, the V.\ M.\ Willaman Endowment, and Penn State ACIS Instrument Team Contract SV4-74018 (issued by the \Chandra \hbox{X-ray} Center, which is operated by the Smithsonian Astrophysical Observatory for and on behalf of NASA under contract NAS8-03060). BL acknowledges financial support from NSFC grants 11991053 and 11673010 and National Key R\&D Program of China grant 2016YFA0400702. The \Chandra ACIS team Guaranteed Time Observations (GTO) utilized were selected by the ACIS Instrument Principal Investigator, Gordon P.\ Garmire, currently of the Huntingdon Institute for \hbox{X-ray} Astronomy, LLC, which is under contract to the Smithsonian Astrophysical Observatory via Contract SV2-82024.

For this research, we have used the Python language along with Astropy\footnote{\url{https://www.astropy.org/}} \citep{astropy2018}, Scipy\footnote{\url{https://www.scipy.org/}} \citep{scipy}, and TOPCAT\footnote{\url{http://www.star.bris.ac.uk/~mbt/topcat/}} \citep{Taylor2005}.

\section*{Data Availability}
The data used in this investigation are available in the article and in its online supplementary material. See Appendix \ref{append:A} for more details.




\bibliographystyle{mnras}
\bibliography{Running_bib} 


\appendix

\section{Quasars in the Full sample}\label{append:A}
Here we present the table containing the \hbox{X-ray} measurements for all of the quasars in the Full sample that have multiple \Chandra observations. A subset of the 27 total columns is reported in Table~\ref{tab:schema}. The full table is available online in machine-readable format.

\begin{table*}
\centering
\caption{Quasars with duplicate \Chandra observations}
\label{tab:schema}

\begin{tabular}{|l|r|r|r|r|r|r|c|}
\hline
  \multicolumn{1}{|c|}{NAME} &
  \multicolumn{1}{c|}{RA} &
  \multicolumn{1}{c|}{DEC} &
  \multicolumn{1}{c|}{$z$} &
  \multicolumn{1}{c|}{ObsID} &
  \multicolumn{1}{c|}{Full\_cts} &
  \multicolumn{1}{c|}{TSTART} &
   \multicolumn{1}{c|}{BAL\_flag} \\
  
  \multicolumn{1}{|c|}{} &
  \multicolumn{1}{c|}{(J2000 deg)} &
  \multicolumn{1}{c|}{(J2000 deg)} &
  \multicolumn{1}{c|}{} &
  \multicolumn{1}{c|}{} &
  \multicolumn{1}{c|}{} &
  \multicolumn{1}{c|}{(s)} &
  \multicolumn{1}{c|}{} \\
  
  \multicolumn{1}{|c|}{(1)} &
  \multicolumn{1}{c|}{(2)} &
  \multicolumn{1}{c|}{(3)} &
  \multicolumn{1}{c|}{(4)} &
  \multicolumn{1}{c|}{(9)} &
  \multicolumn{1}{c|}{(15)} &
  \multicolumn{1}{c|}{(21)} &
  \multicolumn{1}{c|}{(23)} \\
  
\hline
  021000.22$-$100354.2 & 32.5009 & $-$10.0650 & 1.976 & 15666 & 238.225 & 4.905E8 & 0 \\
  021000.22$-$100354.2 & 32.5009 & $-$10.0650 & 1.976 & 15667 & 486.303 & 4.907E8 & 0 \\
  095732.04+024301.7 & 149.3835 & 2.7171 & 0.849 & 15258 & 284.474 & 5.050E8& 0 \\
  095732.04+024301.7 & 149.3835 & 2.7171 & 0.849 & 15259 & 274.178 & 5.072E8& 0 \\
  233722.01+002238.8 & 354.3417 & 0.3774 & 1.382 & 3248 & 9.930 & 1.425E8& 0 \\
  233722.01+002238.8 & 354.3417 & 0.3774 & 1.382 & 11728 & 27.384 & 3.669E8& 0 \\
\hline
\end{tabular}

\begin{flushleft}
\footnotesize{{\it Notes:}
Selected columns from the table of quasars with duplicate \Chandra observations. Quasars without BALs present in the spectrum have a BAL\_flag = 0 whereas quasars that exhibit BALs have BAL\_flag = 1. Appendix \ref{append:A} describes all of the columns in the full table. The complete table is available in machine-readable format.}
\end{flushleft}

\end{table*}

\begin{itemize}

\item[--] Column (1): Object Name.
\item[--] Column (2): J2000 Right Ascension (J2000 degrees).
\item[--] Column (3): J2000 Declination (J2000 degrees).
\item[--] Column (4): Redshift (see \citealt{Richards2015, Paris2018}).
\item[--] Column (5): Galactic column density (cm$^{-2}$; \citealt{Kalberla2005}).
\item[--] Column (6): MJD of the \Chandra observation (days).
\item[--] Column (7): Apparent $i$-band magnitude.
\item[--] Column (8): Reddening in the $i$-band (from the \citealt{Schlafly2011} dust map; subtract columns 7 and 8 to obtain the de-reddened $i$-band magnitude.).
\item[--] Column (9):  \Chandra observation ID.
\item[--] Column (10):  \Chandra off-axis angle (arcmin).

\item[--] Column (11): Full-band effective exposure time (seconds).

\item[--] Column (12): Binomial probability of detection (soft band).
\item[--] Column (13): Binomial probability of detection (hard band).
\item[--] Column (14): Binomial probability of detection (full band).
\item[--] Column (15)--(17): Net counts in the full band; $1\sigma$ upper and lower limits of the net full-band counts.
\item[--] Column (18)--(19): Mean exposure map pixel value of the source and background regions (cm$^2$s).
\item[--] Column (20): Chip-edge flag (0 = good detection; 1 = edge detection).
\item[--] Column (21): \Chandra start time (s).
\item[--] Column (23): Bright cluster flag (0 = no cluster; 1 = cluster).
\item[--] Column (23): BAL flag (0 = no BAL detected; 1 = BAL present).
\item[--] Column (24): Absolute magnitude (corrected to $z=2$; \citealt{Richards2006}).
\item[--] Column (25): Logarithm of the rest-frame monochromatic 2500 \AA\ luminosity (erg s$^{-1}$ Hz$^{-1}$ ).
\item[--] Column (26): Logarithm of the rest-frame monochromatic 2500 \AA\ flux density (erg cm$^{-2}$  s$^{-1}$ Hz$^{-1}$).
\item[--] Column (27): Logarithm of the radio-loudness parameter, $R$.

\end{itemize}

\bsp	
\label{lastpage}
\end{document}